%% file: ms-accepted.tex
\documentclass[twocolumn]{aa} 
\usepackage{graphicx} 
\usepackage{amsbsy} 
\usepackage{wasysym} 
\usepackage{natbib} 
\usepackage{mathtools} 
\usepackage{amsmath} 
\usepackage[dvipsnames]{xcolor} 
\usepackage{newfloat} 
\usepackage[utf8]{inputenc}
\usepackage[colorlinks=True,allcolors=blue]{hyperref} 
\usepackage{multirow} 
\bibpunct{(}{)}{;}{a}{}{,} 

\begin{document} 
 
   \title{SiO maser astrometry of the red transient V838~Monocerotis} 
 
   \author{Gisela N.\ Ortiz-Le\'on \inst{1} 
           \and 
          Karl M.\ Menten\inst{1} 
          \and 
          Tomasz Kami{\' n}ski\inst{2,3} 
          \and 
          Andreas Brunthaler\inst{1} 
          \and 
          Mark J.\ Reid\inst{2} 
          \and 
          Romuald Tylenda\inst{3} 
          } 
 
   \institute{ 
              Max Planck Institut f\"ur Radioastronomie, Auf dem
H\"ugel 69,  D-53121 Bonn, Germany                                      
              \email{gortiz@mpifr-bonn.mpg.de} 
             \and 
             Center for Astrophysics $|$ Harvard \& Smithsonian,
60 Garden Street, Cambridge, MA 02138                                   
             \and 
             Nicolaus Copernicus Astronomical Center, Polish Academy
of Sciences, Rabia{\' n}ska 8, 87-100                                 
             Toru{\' n}} 
 
   \date{Received December 20, 2019} 
 
 
  \abstract{ 
 We present multi-epoch observations with the VLBA of SiO maser emission in the
$v=1$, $J=1-0$ transition at 43~GHz from the remnant of the red nova V838~Mon. 
We model the positions of maser spots to derive a parallax of $0.166\pm0.060$ mas.  
Combining this parallax with other distance information results in a distance of
$5.6\pm0.5$~kpc, which agrees with an independent geometric distance of $6.1\pm0.6$~kpc
from modeling polarimetry images of V838~Mon's light echo. Combining
these results, and including a weakly constraining Gaia DR2 parallax,
yields a best estimate of distance of $5.9\pm0.4$~kpc. 
The maser spots are located close to the peaks of continuum at $\sim$225~GHz
and SiO $J$=5--4 thermal emission detected with ALMA. The proper
motion of V838~Mon confirms its membership in a small open
cluster in the Outer spiral arm of the Milky Way.} 
 
   \maketitle 
%
\section{Introduction}\label{sec:intro} 
 
V838~Mon is a remnant of a so-called red nova, which erupted in January
2002 \citep{MunariDiscovery} and immediately came to prominence to
the astronomical community and the public due to the spectacular
images of a light echo taken with the {\it Hubble Space Telescope}
(HST) \citep{Bond2003}. This eruption probably resulted from a merger
of two ($\simeq8~M_\odot$ and $\simeq0.3~M_\odot$) stars \citep{Tylenda2006},
but other mechanisms have also been proposed.    
\cite{Munari2002IAUC} found a composite optical (post-outburst) spectrum
of V838~Mon consistent with a binary system of a cool giant and a
hotter stellar companion \citep[see also][]{Wagner2002IAU,Crause2003,Kaminski2009}.                                                          
Late in 2002, the cool giant resembled a late-M type star, while the hot
companion appeared to be a B3\,V star. The M-type star is naturally assumed to be the
successor (remnant) of the outbursting star. More recent spectroscopic
observations by \cite{Loebman2015} have classified the cool star
as an L3 supergiant with an effective temperature of $\sim2000-2200$~K.
The separation between the two components was first estimated to be 28~AU based on the assumption that V838 Mon is an eclipsing binary \citep{Munari2007}. This assumption has been proven to be incorrect and a system separation of $\sim$250~AU was found instead \citep{Tylenda2009}.
 
Several scenarios have been proposed to explain the nature of the
2002 eruption (see the extended discussion in \citealt{Tylenda2006}),
although the observational data leave a stellar-merger event as
the most viable explanation. In this scenario, the progenitor of
V838~Mon was likely a triple or higher order system dominated by
two B-type stars (the outbursting star and the B3V companion; \citealt{Tylenda2006}).
A low-mass companion might have entered in a highly eccentric orbit
and finally merged with the primary.                                    
 
\cite{AfsarBond} discovered that V838~Mon belongs to a sparse open
cluster of young ($\lesssim25$~Myr) B-type stars.                       
A diffuse molecular cloud within the echo region was revealed in
CO single-dish radio observations by \cite{Kaminski2011}. This cloud
is thought to be interstellar in nature, consisting of material that
remained after the formation of the cluster to which V838~Mon belongs.  
 
The immediate surroundings of V838~Mon are rich in molecular gas. 
Optical and infrared spectroscopy reveal a multitude of atomic lines
and molecular bands with very complex kinematics \citep{Tylenda2011}.
Part of this neutral cool gas was lost during or just before the
merger, as in other Galactic red novae \citep{Kaminski2018}. Some
of the absorption features of V838~Mon indicate the presence of ongoing
mass loss from the coalesced star \citep{Kaminski2009}.

Searches for maser emission from oxygen-rich molecules (SiO, H$_2$O
and OH) and hydrogen recombination lines have been conducted toward
V838~Mon. \cite{Deguchi2005} made the first detection of  SiO maser
emission ($J=1-0$, $v=1$ and $v=2$)  at 43~GHz with the Nobeyama
45-meter telescope in 2005 (three years after the outburst). \cite{Claussen2007}
reported follow-up SiO maser observations taken with the Very Large
Array (VLA), the Green Bank Telescope (GBT), and the Very Long Baseline
Array (VLBA).  The maser emission observed with the VLA and the GBT
showed variations in flux density superimposed on a rising trend
(reaching a maximum of about 8~Jy) over a $\approx$900-day
timescale, after which the flux started to decline. The VLA
spectra showed one and two features in the $v=2$ and $v=1$ transitions,
respectively, at LSR velocities near 54~km~s$^{-1}$. However, results
from the VLBA observations have not been published.                                    
 
More recently, the same SiO maser lines were observed using the Effelsberg
100m telescope in 2013 October and in 2017 January. It was found
that between the two epochs the $v=2$ line declined from $\approx$0.3~Jy
to $<$0.1~Jy, and the $v=1$ line declined from $\approx$2~Jy to
$\approx$1~Jy. This is consistent with the gradual decline of flux
density which started in early 2006, as reported by \citet{Claussen2007}.
 
SiO $v=1$, $J=2-1$ maser emission at 86~GHz was also detected in
this source with the 45m telescope at Nobeyama \citep{Deguchi2009}
and more recently with the IRAM-30m telescope \citep{Kaminski2018}.
In addition, interferometric observations at millimeter/submillimeter
wavelengths using the Submillimeter Array (SMA) revealed lines of 
CO, SiO, SO, SO$_2$ and H$_2$S covering a very broad velocity range 
($\sim$400~km~s$^{-1}$; \citealt{Kaminski2018}).
These molecular transitions trace thermal emission
from the merger ejecta and are orders of magnitude weaker in peak
intensity than the SiO maser lines.                                     
 
Originally the distance to V838~Mon was uncertain, with estimates
ranging from 6 to 10~kpc \citep{Bond2003,Tylenda2004,Munari2005}.
Based on the geometry of the light echo seen in polarimetric images taken
with the HST, \cite{Sparks2008} derived a distance of $6.1\pm0.6$~kpc. 
This value has been adopted by various authors when deriving the properties
of the remnant.                                                         
 
In this paper, we report new VLBA observations of the 43~GHz SiO
maser transitions toward V838~Mon. The phase-referencing technique
was used to provide accurate astrometry. The VLBA observations are
presented in Section \ref{sec:obs} and their analysis in Section
\ref{sec:detections}. The astrometric parameters and the distance
and proper motion to the source are derived in Section \ref{sec:astro}.
Over the course of the VLBA observations, the Gaia mission released
its second catalog (DR2; \citealt{Gaia2018,Lindegren2018}). We thus
also compare our astrometry with that provided by Gaia. We discuss
the parallax, position, and proper motions of V838~Mon in Section
\ref{sec:discuss}, as well as SiO thermal emission detected in new
ALMA observations. Finally, we present our conclusions in Section
\ref{sec:concluss}.                                                     
 
\section{Observations and data reduction}\label{sec:obs} 
 
\subsection{VLBA observations} 
 
V838~Mon was observed with the VLBA for a total of 7 epochs between
2017 October and 2019 March (Table \ref{tab:obs}).  The observations
were taken at 43~GHz with 4 intermediate frequency (IF) bands of
16~MHz bandwidth. Except for the first epoch, two of these IFs were
centered at the $v=1$, $J=1-0$ and $v=2$, $J=1-0$ SiO maser transitions,
at 43122.080~MHz and 42820.582~MHz, respectively \citep{Mueller2013},
and correlated with a channel spacings of 31.25~KHz (0.2~km~s$^{-1}$).
The first epoch only covered the $v=1$, $J=1-0$ transition. The quasars
J0709--0255 (RA=07:09:45.0546, Dec=--02:55:17.496) and J0656--0323
(RA=06:56:11.1206, Dec=--03:23:06.782), in J2000 coordinates,
were observed as phase reference calibrators.  Observations
consisted of alternate scans on the target and one phase-reference calibrator,
switching sources every $\approx$20 seconds.                            
Blocks of about 30 minutes spent on calibrators distributed over
a wide range of elevations were observed at 23.7~GHz every $\approx$2
hours during each 7-hr observing run. These scans (the so called
geodetic-like blocks) were used to estimate multi-band delays,
which are predominantly introduced by residual tropospheric delays 
and clock errors.                    
 
Data calibration was performed using the Astronomical Imaging System
(AIPS) \citep{Greisen2003}, using a ParselTongue scripting interface
\citep{Kettenis2006}.  We used procedures that have been widely applied to
perform high-frequency astrometry  as part of the Bar and Spiral
Structure Legacy (BeSSeL) Survey at 22~GHz \citep{Reid2009}. 
Given the 300~MHz frequency separation between the $v=2$ and $v=1$
transitions, we derived independent delay and rate solutions for each IF
band instead of combining them. Also, since the detected maser emission
in V838~Mon is relatively weak (<1~Jy; Section \ref{sec:detections}),
the fringe-fitting solutions were derived from the scans on a phase-reference
calibrator and then applied to the target. Phase solutions from scans
on J0656--0323 were found to have higher SNR than the solutions from
J0709--0255. We therefore only used the data from J0656--0323 for
the phase-referencing process.                                          
 
After calibrating the visibility data, individual channel maps of
$2048\times2048$ pixels of $60~\mu$as were constructed using the AIPS 
task IMAGR. We imaged all spectral channels covering
a range in LSR velocity of -1 to 110~km~s$^{-1}$.                         
 
\subsection{ALMA observations} 
V838\,Mon was observed by ALMA in band 6 with four spectral ranges:
215.56--217.44, 219.06--220.94, 230.11--231.99, and 232.06--233.94~GHz
(PI: T. Kami\'nski). The observations took place on 20 June 2019
with the longest ALMA baselines of 15.3\,km, which resulted in an
angular resolution of 22\,mas (when imaged with Briggs weighting
and with the robust parameter of 0.5). Visibilities were calibrated
with the default calibration scripts delivered by the observatory
and were imaged in CASA \citep{casa}. The ALMA data will be presented
and analyzed in depth in a dedicated paper.                             
Here, we present only maps of dust continuum emission and the
SiO $v=0$, $J$=5--4 line in both of which the source is clearly resolved.

\section{Results}\label{sec:detections} 
 
\begin{figure*}[!bht] 
\begin{center} 
 {\includegraphics[width=0.5\textwidth,angle=0]{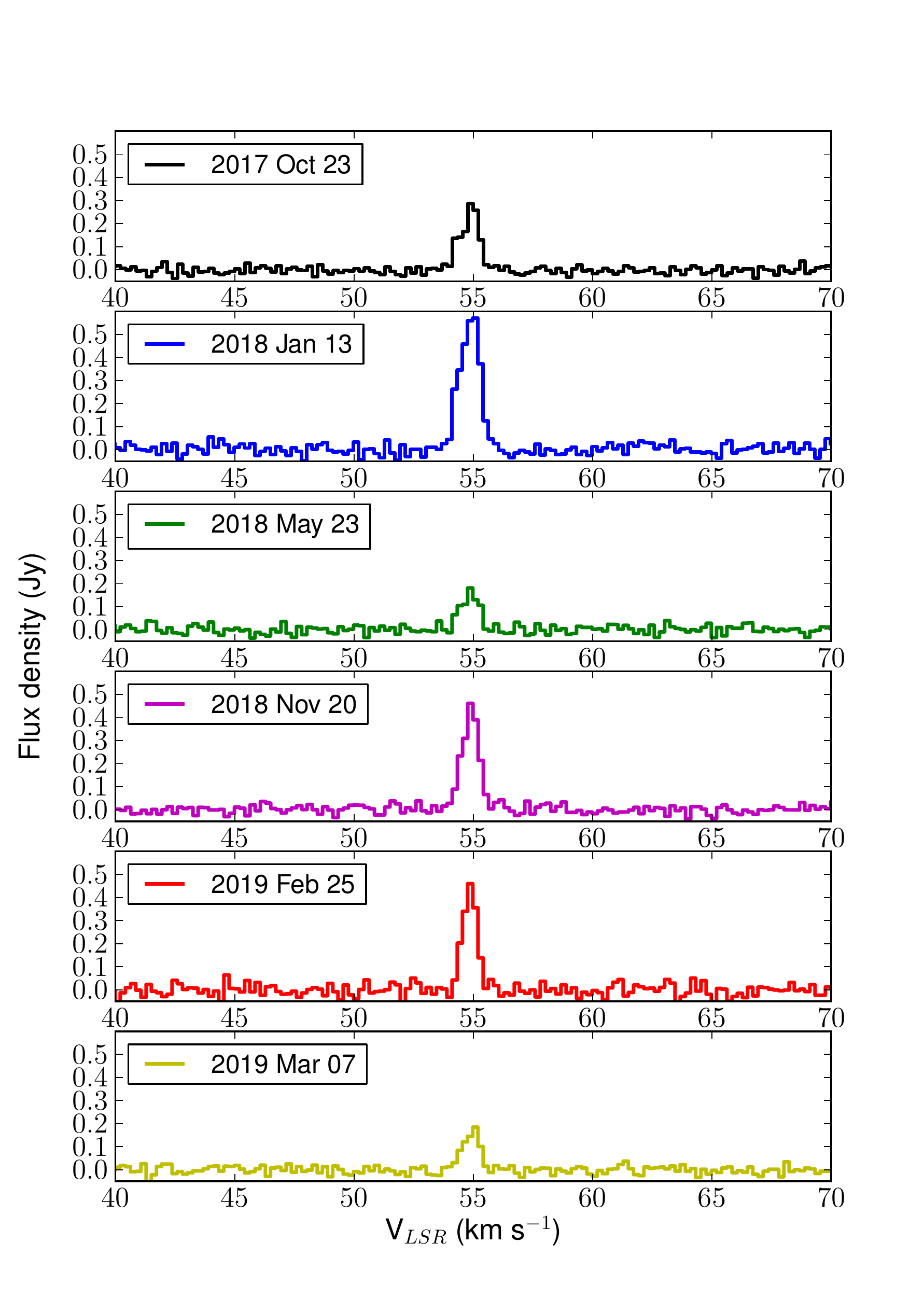}} 
 \end{center} 
\caption{Integrated flux density spectra of the SiO maser ($v=1$,
$J=1-0$) detected with the VLBA at the epoch indicated by the label
in each panel. }                                                        
\label{fig:spectra} 
\end{figure*}

We detected maser emission from the $v=1$, $J=1-0$ transition 
at 6 epochs at LSR velocities
from $\approx 54$ to 58~km~s$^{-1}$ with a peak near 55~km~s$^{-1}$
(Figures \ref{fig:spectra} and \ref{fig:e1}). Poor observing conditions
on August 27, 2018 degraded the sensitivity of the array, which
resulted in a non-detection of the $v=1$ line. Integrated
flux density spectra of the maser emission are shown in Figure \ref{fig:spectra}.
 
{The $v=2$, $J=1-0$ was marginally detected only on January 13, 2018,
with an integrated flux density of $0.10\pm0.03$~Jy at 55.3~km~s$^{-1}$.
This line was not detected in the last 5 epochs}.                    
 
The peak flux density of the $v=1$ transition, detected with Effelsberg
in 2017 January, was 1.2~Jy, i.e.\ almost two times stronger than
we see with the VLBA. This variation in the maser intensity
is more pronounced than observed before (\citealt{Claussen2007}).
In order to investigate this in more detail, we performed a phase 
self-calibration, using the  velocity channel with the peak flux.
The flux density measured in the resulting maps from all epochs 
after three iterations of self-calibration was $0.7-1.2$~Jy. Thus,
we are detecting lower flux density in the non-self-calibrated cubes owing to
of residual phase errors. Other contributions to the discrepancy
in flux density when compared to Effelsberg may come from large-scale
emission that is resolved out by the VLBA (see \citealt{Claussen2007})
and the overall decline in strength that started in early 2006. Since
the absolute positions are lost by self-calibrating
the data, we do not use the self-calibrated images constructed
for this analysis. We note that no additional maser components were
found, which suggests those may be too weak to be detected with the
current sensitivity of our observations.                                
 
We detected three maser features in total, where a ``feature'' refers
to emission observed in contiguous velocity channels at nearly
the same position.                                             
Channel maps using the entire $uv$ range are presented in Figure
\ref{fig:e1} for the strongest maser feature. 
Unlike \cite{Claussen2007}, who did not resolve emission in
the peak of the line, our new VLBA maps appear to show some
resolution.                                                
Figure \ref{fig:e1b} shows the emission from the second strongest
feature, which is detected at low SNR and in only four epochs at
$\approx$7~mas northeast of the strongest feature. This feature
is blueshifted by $\approx$0.9~km~s$^{-1}$ relative to the main feature.
Last, a third feature is redshifted ($v_{\rm LSR}\approx58$~km~s$^{-1}$)
and detected only in the first and second epochs at $\approx$1.5~mas
eastward of the strongest feature (Figure \ref{fig:spot3}). Figure
\ref{fig:spots} shows the spatial location of the three maser features.

\section{Astrometry}\label{sec:astro} 
 
In order to derive astrometric parameters for V838~Mon, we first selected
two contiguous velocity channels with the strongest SiO
emission (maser channels at 55.1 and 54.9 km~s$^{-1}$). 
The emission in these channels is resolved
in most epochs, so in order to minimize the effects of extended
emission we imaged the data with a minimum $uv$ length of 100 or 150~M$\lambda$
(the limit of 100~M$\lambda$ was chosen in the epochs where the maser
emission is weak, since omitting the more of the data resulted
in non detections). We then fitted a 2D Gaussian to
the brightness distribution using the AIPS task JMFIT. The
positions  measured in the maps, derived with
J0656--0323 as the reference source, are given in Table \ref{tab:jmfit}.
Then, we simultaneously fit the positions of the two maser channels,
solving for a single parallax ($\varpi$), but allowing for different proper
motions $(\mu_{\alpha}\cos\delta,\mu_{\delta})$ and position offsets 
$(\alpha_{0},\delta_{0})$ at the reference epoch equal to
the mean of the observed epochs.                                        
We should note that the
B3V companion is between 28 and 250~au from the primary, which implies an orbital
period in the range 40 to 1000 yr (assuming a circular orbit). Thus, we do not
expect that the orbital motion has a measurable
effect on the motion of the primary over the 1.4 years of our observations.
Figure \ref{fig:fit} shows the best fit for
the resulting astrometric elements given in Table \ref{tab:astro}.
Our fitting routine computes additional systematic errors to be added
in quadrature to the statistical source position errors provided
by JMFIT. These errors are estimated numerically so as to make the
reduced $\chi^2\approx1$ for each coordinate. The resulting best
fit parallax is                                                         
 
\begin{eqnarray} 
\varpi & = & 0.166  \pm 0.060~{\rm mas}~~~~, 
\end{eqnarray} 
 
\noindent where we have multiplied the parallax uncertainty by $\sqrt{2}$
to account for possible systematic errors, which would be fully correlated
for the two maser channels. Fitting each spot separately gives similar
results for all astrometric parameters, but with slightly larger
uncertainties.                                                          
 
We note that the fractional parallax error is significantly large,
thus, we should take particular care when estimating the distance
and its uncertainty from the parallax. Since simply inverting the parallax
yields a poor approximation of the distance, we used Bayesian inference.
For a truncated uniform prior, an unnormalized posterior probability
density function (PDF) given by \citep{Bailer-Jones2015} is as follows:            
 
\begin{equation} 
P^{*}(r~\vert~\varpi, \sigma_\varpi) = 
\begin{cases} 
      \frac{1}{ r_{\rm lim}} \frac{1}{\sqrt{2\pi}\sigma_\varpi} \exp{
\Bigl[ - \frac{1}{2\sigma_\varpi^2} \left( \varpi - \frac{1}{r} \right)
^2  } \Bigr] & {\rm if}~r<0\leq r_{\rm lim} \\                          
       0 & \rm{otherwise}.  \\ 
   \end{cases} 
   \label{eq:posterior} 
   \end{equation} 
 
\noindent
This astrometric-based posterior PDF for
$r_{\rm lim}=20$~kpc is shown in the right panel
of Figure \ref{fig:pdf-all}.
                                                   
Finally, we apply the parallax-based distance estimator of \cite{Reid2016,Reid2019} 
for sources associated with spiral arms in the Milky Way in order  
to derive a combined PDF.  This estimator combines other distance information including
spiral arm assignment (SA), kinematic distances based on radial velocity (KD) and two 
components of proper motion (PM(long) and PM(lat)): 
 
\begin{center} 
\begin{equation} 
{\rm Prob}(d) \propto 
{\rm Prob}_{\rm SA}(d) \times {\rm Prob}_{\rm KD}(d) \times {\rm
Prob}_{\rm PM(long)}(d) \times {\rm Prob}_{\rm PM(lat)}(d)~~~~.                         
   \label{eq:reid} 
\end{equation} 
\end{center} 
 
\noindent 
The left panel in Figure \ref{fig:pdf-all} shows the individual parameters
of the parallax-based distance estimator, as well as
the combined PDF.  We have used $v_{\rm LSR}=55\pm5$~km~s$^{-1}$ and
the average of the proper motions of the two maser spots,  $\mu_\alpha
\cos \delta=-0.465\pm 0.056$~mas~yr$^{-1}$ and $\mu_\delta=0.791\pm0.065$~mas~yr$^{-1}$.
The source is most likely associated with the Outer spiral arm at
a distance of $5.5$~kpc (as previously suggested by \citealt{Kaminski2011} and
\citealt{Quiroga2019}).

For the second strongest spot, we were able to only estimate its
proper motion using the positions fitted to the brightness distribution
at $v_{\rm LSR}$ of $54.2$~km~s$^{-1}$ (listed in 
Table \ref{tab:astro} and shown in Figure \ref{fig:spots}).
 
We have searched the Gaia-DR2 catalog and found a star (2MASS J07040482-0350506)
within a search radius of 1~arcsec of the SiO maser position. When propagating
the Gaia position at J2015.5 to the VLBA observing epochs, we found
that this source has a separation of only 2.4~mas from the 
VLBA position of the strongest maser.  This star has a DR2 parallax 
of $-0.001\pm0.105$~mas.
For comparison with our VLBA distance PDF, we have constructed the
Gaia-DR2 distance PDF using a truncated uniform prior as shown in the 
right panel of Figure \ref{fig:pdf-all}).
Similarly, we show the posterior PDF that corresponds to the light-echo
distance measurement by \cite{Sparks2008}. These PDFs can be combined
with the distance PDF given by equation (\ref{eq:reid}) to provide a
more robust distance estimate than either alone. The combination
of equation (\ref{eq:reid}) with the VLBA astrometric-based PDF yields
$d=5.5\pm0.6$~kpc, while the combination of all PDFs yields $d=5.9\pm0.4$~kpc,
where the distance and its uncertainty have been obtained by fitting
a Gaussian model to the combined PDF and estimating its peak probability
density, center and width. Table \ref{tab:prlx} summarizes all the
distance results discussed above. We include the Gaia result for 
completeness even though it is highly uncertain and contributes little 
to the final distance estimate. Our study provides  
an example that Gaia DR2 parallaxes for distant supergiants should be 
treated with caution.                                   
 
As described in the introduction, V838~Mon is a binary system consisting
of a cool giant (the eruptive component) and the B3V companion. An
effective temperature of $T_{\rm eff}\sim2000-3000$~K  was measured
for the merged star from optical and IR spectroscopy 
in 2008--2009 \citep{Tylenda2011,Loebman2015}. The B3V companion
has been attenuated by the V838~Mon outburst ejecta \citep{Goranskij2008},
which explains why it is not detected by Gaia. It is important to
note that the astrometric parameters measured by Gaia toward V838~Mon
may be affected by extinction from the dusty region around V838~Mon.
The DR2 quality factors of the solution suggest that the source is
astrometrically well-behaved; however, the relative errors on the
astrometric parameters are quite large and the parallax is negative.
The Gaia proper motion measurement strongly disagrees with that measured
by the VLBA, as well as the Gaia proper motions of other stars in the V838~Mon
cluster (see Section \ref{sec:cluster}).

\bigskip 
 
\section{Discussion}\label{sec:discuss} 
 
\subsection{The cluster of V838 Mon}\label{sec:cluster} 
 
Gaia DR2 parallaxes and proper motions of the three stars that
are considered to be members of the open cluster of V838 Mon  \citep{AfsarBond},
although highly uncertain, are similar to those derived
by our VLBA measurements.                                                              
These stars are spectroscopically identified as B-type main-sequence
stars. Their individual parallaxes are between 0.17 and 0.24 mas and have
a weighted-mean parallax of 0.18$\pm$0.10 mas, after correcting for
a Gaia DR2 parallax zero-point shift of $-0.029$~mas\footnote{We note that
larger values for the zero-point offset have been reported in the
literature \citep[e.g.\ $-0.08$~mas  by][]{Stassun2018}. Here and
throughout the paper we use the value of $-0.029$~mas derived by
the Gaia-DR2 team \citep{Lindegren2018}.}, which agrees with our 
VLBA parallax of V838 Mon. Their mean proper motion is  --0.49$\pm$0.11
and 0.54$\pm$0.11~mas~yr$^{-1}$ in the easterly and northerly directions, 
respectively\footnote{We
have added quadratically systematic errors of 0.1~mas for parallax
and 0.1~mas~yr$^{-1}$ for proper motions \citep{Luri2018}.}. Note
that the mean proper motion in declination of the SiO masers is consistent
with the mean cluster motion within $\sim$2$\sigma$ uncertainty,
whereas the agreement in the parallaxes and RA proper motion is even
better.                                                                 
 
We note that the Gaia DR2 effective temperatures for the three
B-type stars with known visual spectra \citep{AfsarBond} appear
significantly low, most
likely owing to a high extinction ($E_{B-V} \approx 0.8$)
towards the cluster not taken into account in
the Gaia catalog \citep{GaiaTeff}. This puts into question 
the Gaia effective temperature listed at 3342\,K  for V838 Mon itself, 
although this value is close, within $\mathbf{3\sigma}$, to that expected \citep{Loebman2015}.

We looked for other potential members of the cluster by searching
for sources with similar parallaxes and proper motions as
V838 Mon and the three cluster members. Within a radius
of 2\farcm3 of V838 Mon, we find only one candidate in the Gaia DR2 catalog, star 
3107789583219504000, which is 6\arcsec\ east of V838 Mon. The Gaia parallax  for this
star (0.16$\pm$0.14~mas) is highly uncertain, but consistent with
a distance beyond 3~kpc.                                                
Its location and proper motion are shown in Fig.\,\ref{fig-GaiaCluster}.
Its angular distance from V838 Mon is the smallest among the four field
stars. It is also fainter than the three B-type stars which may be
indicative of a spectral type later than B6. In the maps of \citet{W2003},
this star is numbered 4 and has a slightly smaller interstellar linear
polarization degree and a slightly different orientation of the polarization
plane than the three confirmed cluster members. However, given the
patchy morphology of the echo material and the patchy dust distribution
towards the V838 Mon cluster \citep{Sparks2008,TK2012}, the difference
does not exclude that we have identified a genuine fourth member
of the cluster (not counting the stars in V838 Mon itself).

Given that the considered merger of V838 Mon took place in a triple
or higher system, it is worth considering whether any star (including
V838 Mon itself) could have been ``kicked out'' (or dynamically ejected)
off the system \citep[see e.g.][]{Perets}. This would probably be
manifested by proper motions significantly different than that of
the cluster. This is not observed for V838 Mon at the current accuracy
of the VLBA measurements. Indeed, after subtracting the mean proper motion of
the cluster from the VLBA proper motions, we obtain that V838 Mon  has a 
transverse velocity relative to the cluster of $7\pm5$~km~s$^{-1}$. 
The radial component is unknown. This low velocity implies a small displacement of 
V838~Mon as a result of the ``kick''. If we assume that the ejection took place 
in 2002, we obtain a displacement of only $4\pm2$~mas or $25\pm12$~au.  
The low velocity is inconsistent with the ejection scenario. 
However, constraints on the motion of the now dust-embedded physical companion of V838 Mon 
are very poor \citep{Kaminski2009}, thus we cannot completely rule out that this
component has been dynamically ejected.       
 
\subsection{Location of the stellar photosphere in VLBA maps} 
Optical observations, the photospheric size \citep{Chesneau2014},
and the presence of an SiO maser make V838 Mon very similar to red
supergiants with highest mass-loss rates. SiO masers observed in
AGB stars, and Mira stars in particular, often take the form of a ring
surrounding the star. In such cases, it is relatively straightforward
to identify the location of the stellar photosphere, even if it is
not directly detected in the same observations \citep[see, e.g.,][]{Cotton2008}.
This task is more difficult for red supergiants whose mass loss may
be very inhomogeneous and the distribution of SiO maser spots may
be very erratic. VY\,CMa is the best known example of such a complexity
\citep{Zhang}. In our maps of the maser emission in V838 Mon we probably
detect only the three brightest SiO maser spots. Assuming that their
distribution is as complex as in red supergiants, we are unable to
identify the location of the stellar photosphere in the radio maps.
Given that V838~Mon is an unresolved binary for Gaia, we are also
unable to associate the Gaia position to the V838 Mon merger product
or the companion. 
The stellar remnant of the component of V838~Mon that erupted 
in 2002 remains the brightest radiation source of the system. Part of its energy 
is generated by continuing contraction and thus it has a higher bolometric 
luminosity than the B3V companion. The stellar radiation heats the surrounding 
dust expelled during the eruption producing emission at millimeter wavelengths. 
The emission is brightest closest to the contracting luminous star. 
Thus, we can take the position of the millimeter-wave continuum 
peak as the stellar (photosphere) position. We see in Figure \ref{fig:spots} 
that this peak is located, within the modest accuracy of the ALMA position measurements 
($\approx\pm$3\,mas), between the two brightest maser spots.

\cite{Chesneau2014} observed the mid-IR (dust) emission around V838~Mon
with the Very Large Telescope Interferometer array and derived a major axis
size of 25~mas at 8~$\mu$m and 70~mas at 13~$\mu$m (147 and 412~au at 5.9~kpc, 
respectively). The dust emission is flatter at 13~$\mu$m and has
a position angle of about $-10^{\rm o}$.  
This dusty structure, interpreted as a disk,
is believed to have formed after the outburst in 2002 as a result
of the merger event. For the central star, \cite{Chesneau2014} derived
a diameter of $1.15\pm0.20$~mas ($\approx$6.8~au at 5.9~kpc). The
SiO maser spots are located within the extended dusty structure around
V838~Mon at $\approx$5~mas ($\approx$30~au) from the stellar position
(as measured in the ALMA continuum map), i.e.\ at $\approx$9 stellar
radii but within the dusty structure identified by \cite{Chesneau2014}.
This supports the resemblance of V838~Mon with AGB stars, where the SiO masers
are found in the immediate vicinity of the stellar photosphere \citep{Cotton2008}.

\subsection{Extended emission detected with ALMA} 
 
The velocity-integrated intensity map of the SiO thermal emission
detected by ALMA is shown in Figure \ref{fig:alma}. This corresponds
to the SiO~$v=0$, $J=5-4$ transition at 217~GHz, which shows a very
broad profile with a FWHM of 225~km~s$^{-1}$ for the entire emission
region. The emission has an extended structure with a beam-deconvolved
size of $148\times125~(\pm30)$~mas. The major axis of this structure
is aligned at a position angle of $61\pm7^{\rm o}$. Continuum emission
contours at 225~GHz are overlaid in Figure \ref{fig:alma}.              
The continuum has a peak flux density of 0.43~mJy~beam$^{-1}$ and
an elongated morphology in a direction almost perpendicular to the
major axis of the SiO thermal emission. The extent of the continuum
structure is about 150~mas. Gaussian fits yield the continuum peak
at RA=07:04:04.821535 ($\pm$0.17\,mas) and Dec=--03:50:50.629426
($\pm$0.18\,mas) and the extended SiO emission peaks at RA=07:04:04.822
($\pm$5\,mas) and Dec=--03:50:50.638 ($\pm$5\,mas) (ICRS), where
the errors express only formal uncertainties of the centroiding procedure.
The astrometric accuracy of the ALMA observations is $\approx$3\,mas.
\footnote{\url{https://help.almascience.org/index.php?/Knowledgebase/Article/View/319/6/what-is-the-astrometric-accuracy-of-alma}}
The position of the strongest SiO maser spot agrees within $2\sigma$
with the peaks of continuum and thermal SiO emission (see Figure
\ref{fig:spots}).                                                       
 
As pointed out by \cite{Kaminski2018}, the millimeter emission observed
with ALMA in this and other red novae is produced by material from
the merger ejecta. Then, it is worth noting that the elongated morphology
resembles that of a jet. The formation of jets during a merger event
has been discussed e.g.\ in \cite{Kashi2016}. An elongated morphology
of the merger ejecta is also consistent with mass loss in the orbital
plane of the merging stars \citep[e.g.][]{Pejcha2016}. This requires
the orbit to be almost edge-on. However, the inclination of the orbit,
as well as the rest of the orbital parameters, has not been yet constrained.
The morphology of the millimeter continuum, which traces dust, is
different than that of the SiO, which is tracing the outflow. However,
other lines like $^{12}$CO, $^{13}$CO, AlOH, SO$_2$, and H$_2$S show
emission spatially coincident with the continuum. The thermal submillimeter
emission of V838 Mon will be analyzed in a dedicated study.

\section{Conclusions}\label{sec:concluss} 
 
Multi-epoch VLBA observations of the SiO $v$=1, 2 $J$=1--0 maser
transitions at 43~GHz have been obtained toward the red transient
V838~Mon. Three maser spots from the $v$=1 line were firmly detected,
while the $v$=2 line was tentatively detected in only one epoch.
We modeled the motion of maser spots  which were detected in two
contiguous channels at 6 epochs to fit the astrometric parameters
of the source. The resulting best-fit parallax is $0.166\pm0.060$~mas.
Using Bayesian inference, we combined our VLBA parallax with other
distance information
to arrive at an improved distance of $5.9\pm0.4$~kpc. \cite{Quiroga2019}
and \cite{Reid2019} considered whether V838~Mon is in the Perseus
arm, the Outer arm, or the interam region, and favored the Outer
arm. The improved distance of $5.9\pm0.4$~kpc indicates that it is
further than 5~kpc which favors it being in the Outer arm.

Proper motions of field stars in the vicinity of V838~Mon were taken
from Gaia-DR2. The mean proper motion of the three known members
of the V838~Mon open cluster is consistent within $2\sigma$ with
the VLBA proper motions, confirming its membership. A possible new
member of the cluster was also identified.                              
 
Finally, observations taken with ALMA of SiO thermal emission at
217~GHz revealed an extended outflow with a size of $148\times125$~mas
($870\times$740~au). The major axis of this structure is perpendicular
to 225~GHz continuum emission, which shows an elongated and complex
structure.                                                              
 

\input{observations}

\input{vlba_positions} 
\input{vlba_astro} 
\input{parallaxes_all} 
 

\begin{figure*}[!bht] 
\begin{center} 
 {\includegraphics[width=0.8\textwidth,angle=0]{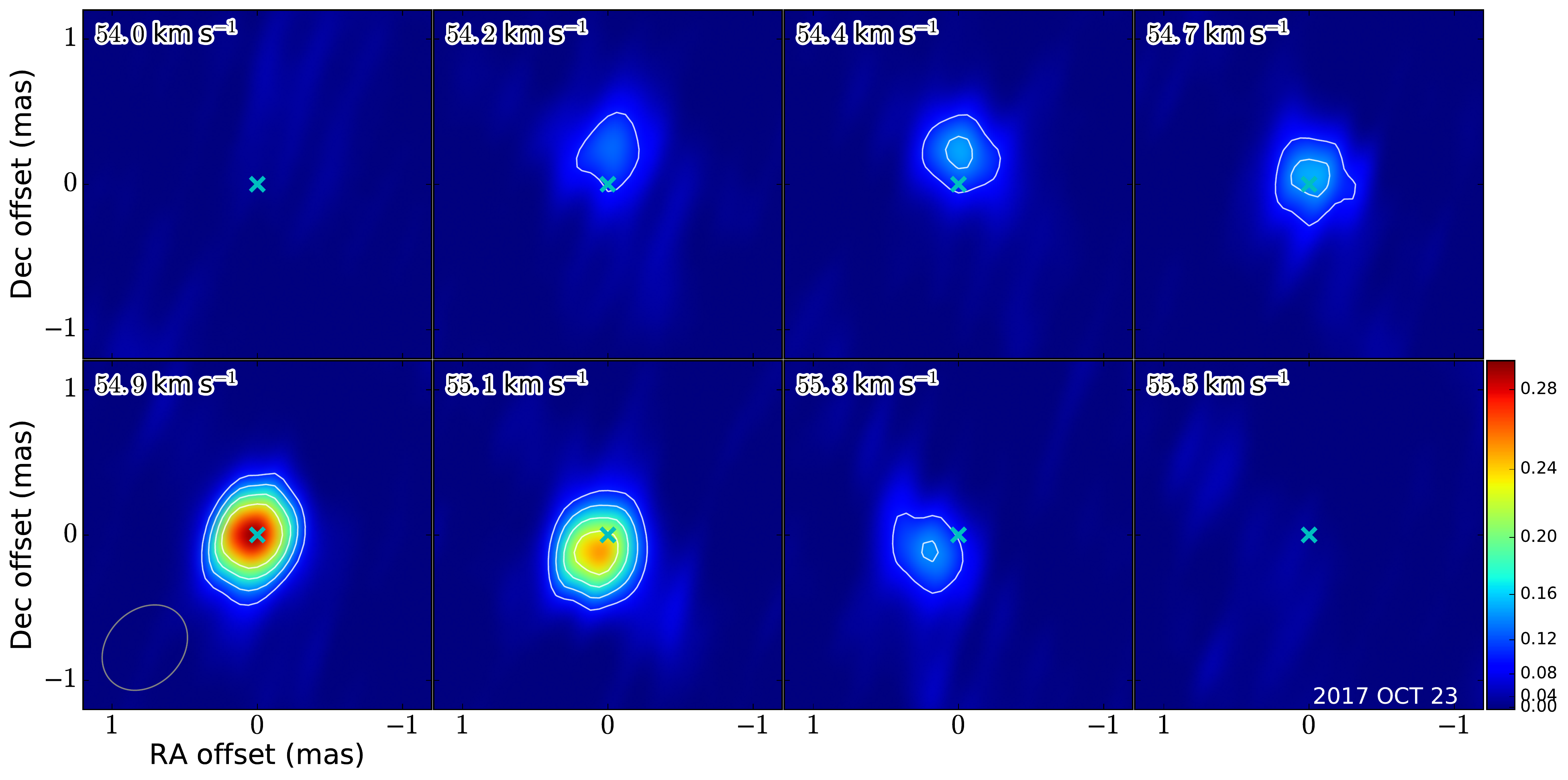}} 
 \end{center} 
\caption{Channel velocity map of the SiO maser ($v=1$, $J=1-0$) emission
for the epoch as indicated by the label in the last panel. The contour
levels are at $5\sigma$, $7\sigma$, $9\sigma$, and $11\sigma$, where
$\sigma$~Jy/beam is the rms noise measured in the strongest channel
of the image cube (Table \ref{tab:obs}). Position offsets are relative
to the the position of the strongest pixel in the channel that has
the maximum peak flux (cyan cross mark).                                
} 
\label{fig:e1} 
\end{figure*} 
 
\setcounter{figure}{0} 
\renewcommand{\thefigure}{2} 
 
\begin{figure*}[!bht] 
\begin{center} 
 {\includegraphics[width=0.8\textwidth,angle=0]{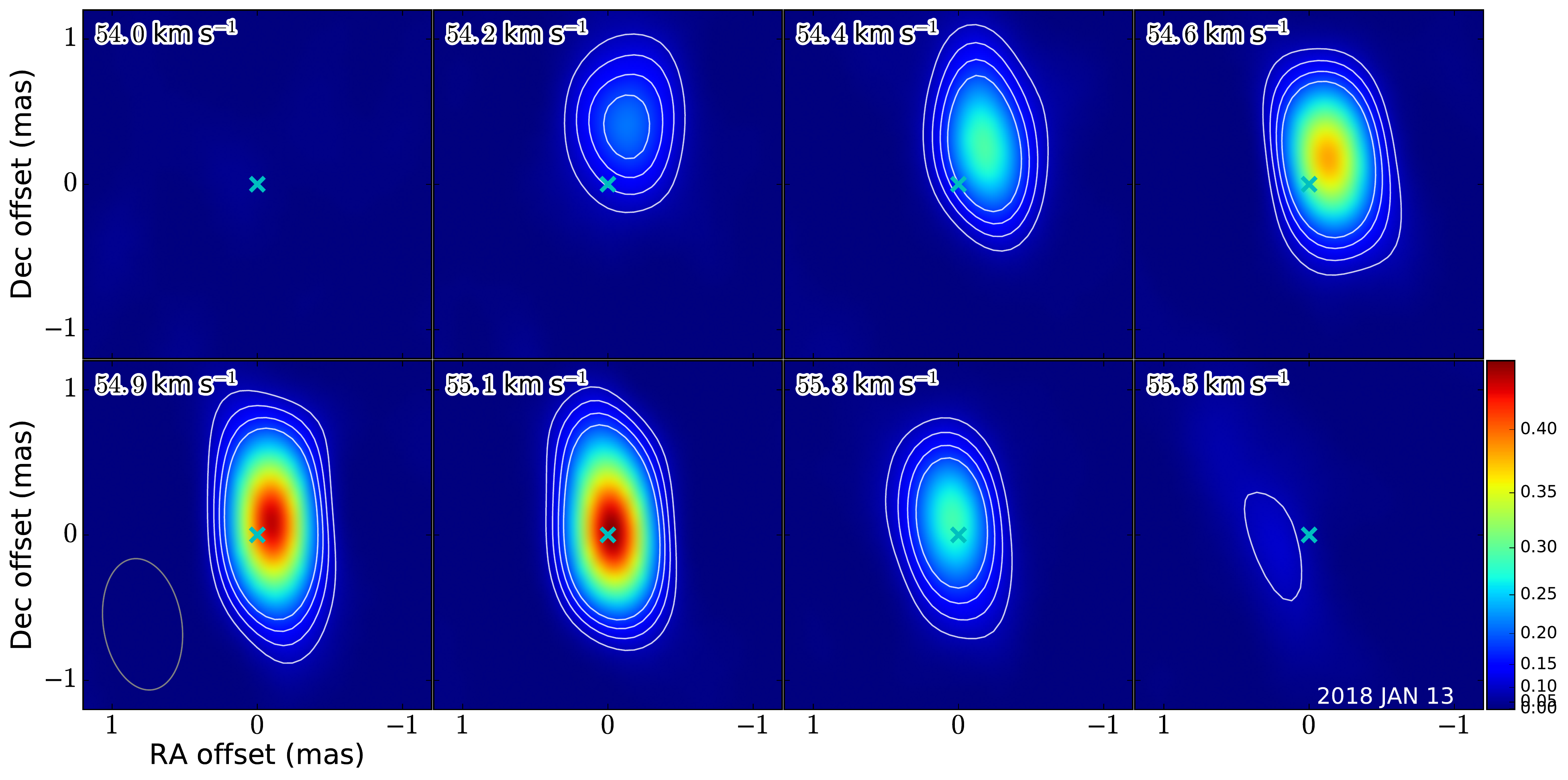}} 
 \end{center} 
\caption{\it Continued. } 
\end{figure*} 
 
\setcounter{figure}{0} 
\renewcommand{\thefigure}{2} 
 
\begin{figure*}[!bht] 
\begin{center} 
 {\includegraphics[width=0.8\textwidth,angle=0]{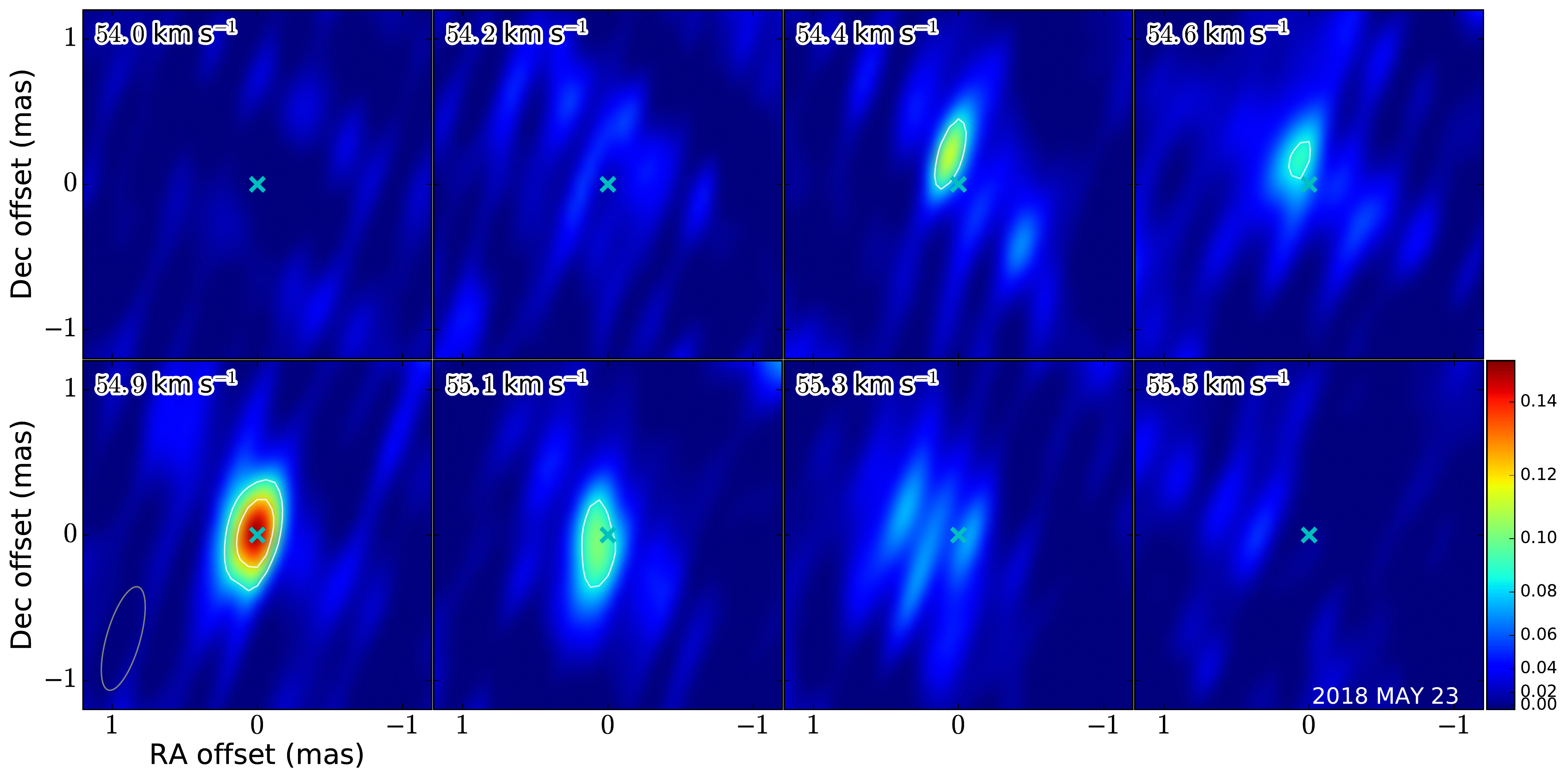}} 
 \end{center} 
\caption{\it Continued. } 
\end{figure*} 
 
\setcounter{figure}{0} 
\renewcommand{\thefigure}{2} 
 
\begin{figure*}[!bht] 
\begin{center} 
 {\includegraphics[width=0.8\textwidth,angle=0]{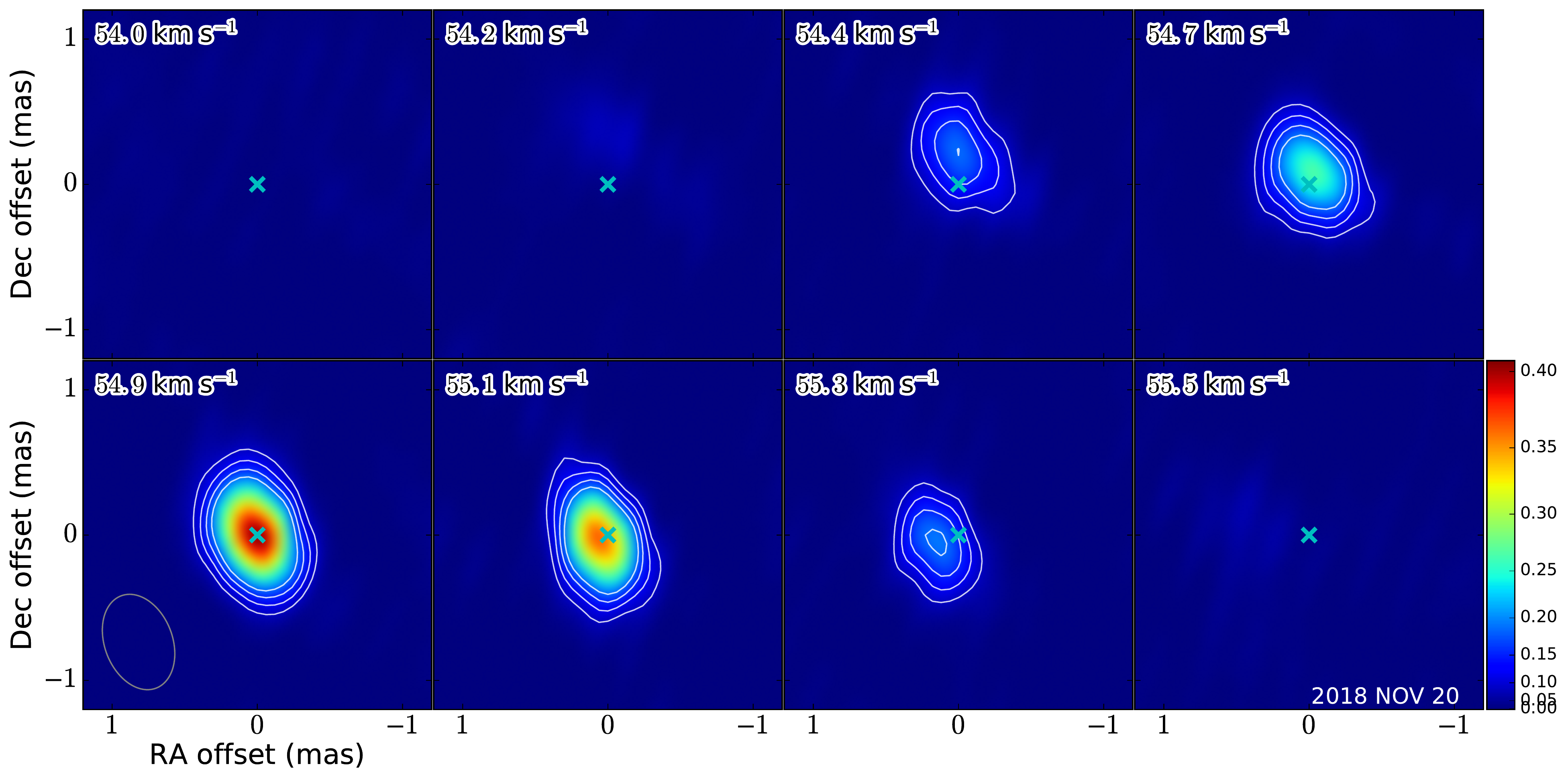}} 
 \end{center} 
\caption{\it Continued. } 
\end{figure*} 
 
\setcounter{figure}{0} 
\renewcommand{\thefigure}{2} 
 
\begin{figure*}[!bht] 
\begin{center} 
 {\includegraphics[width=0.8\textwidth,angle=0]{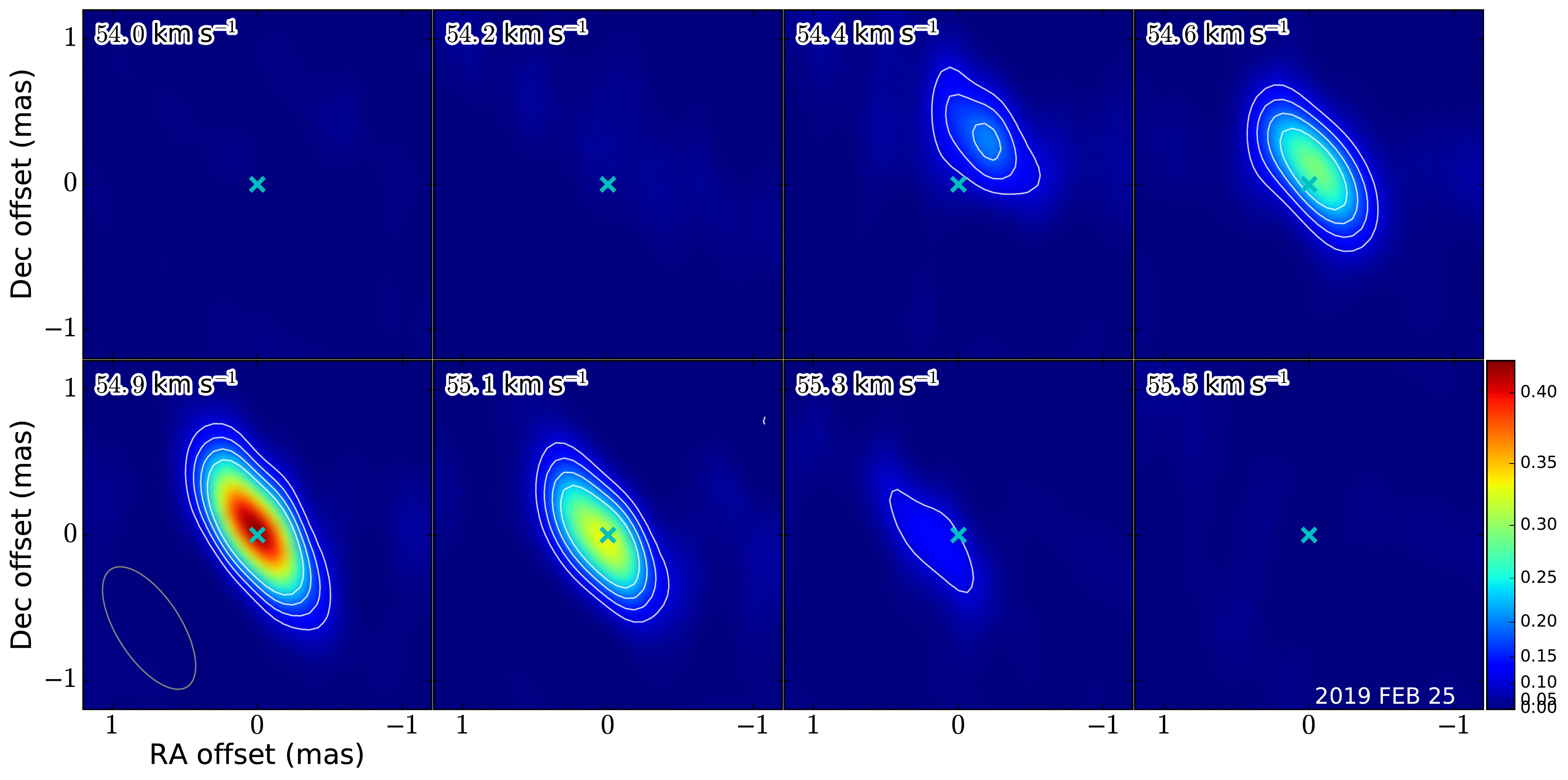}} 
 \end{center} 
\caption{\it Continued. } 
\end{figure*} 
 
\setcounter{figure}{0} 
\renewcommand{\thefigure}{2} 
 
\begin{figure*}[!bht] 
\begin{center} 
 {\includegraphics[width=0.8\textwidth,angle=0]{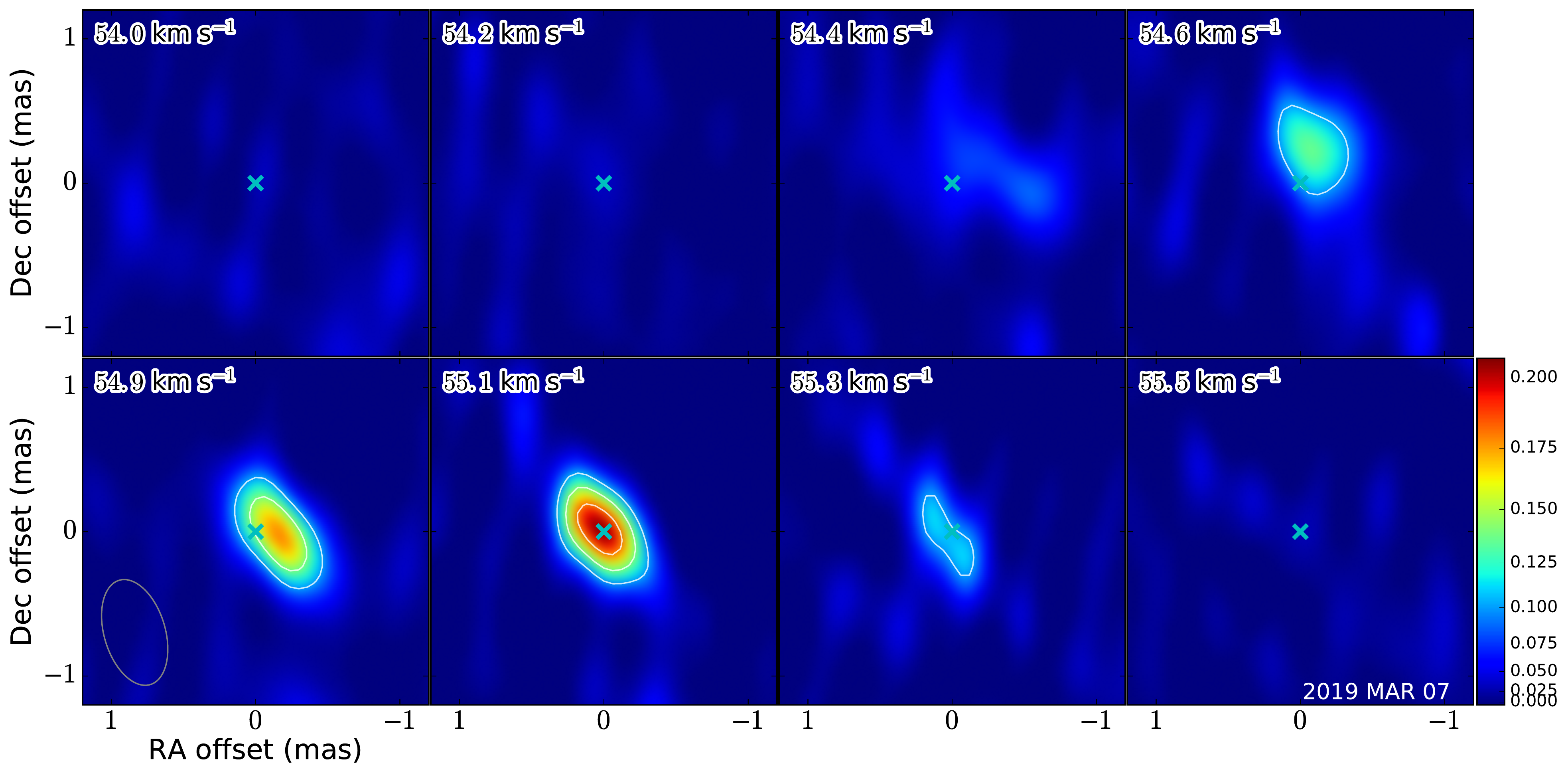}} 
 \end{center} 
\caption{\it Continued.} 
\end{figure*} 
 
\setcounter{figure}{2} 
\renewcommand{\thefigure}{\arabic{figure}} 
 
\begin{figure*}[!bht] 
\begin{center} 
 {\includegraphics[width=0.8\textwidth,angle=0]{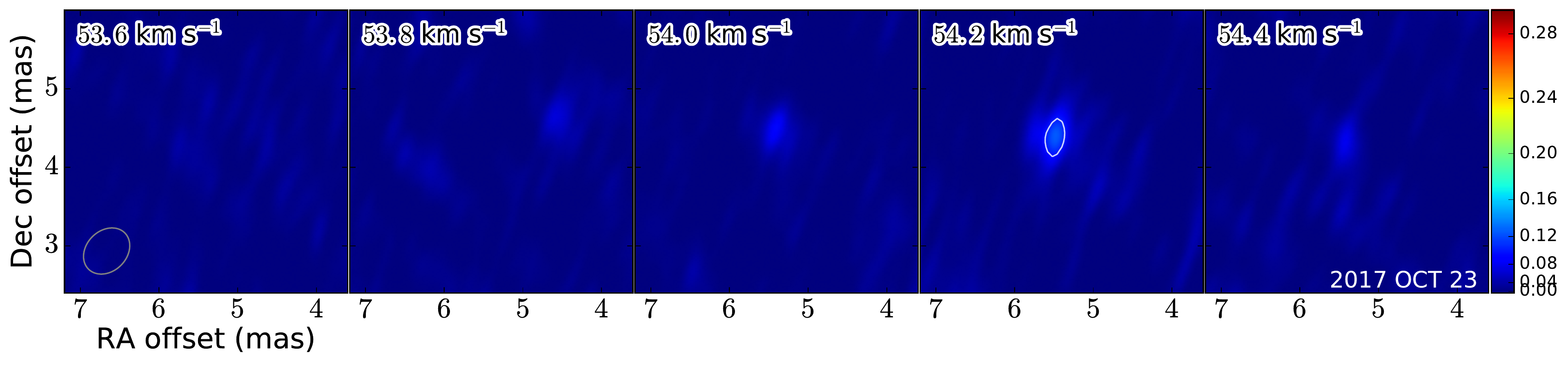}} 
 {\includegraphics[width=0.8\textwidth,angle=0]{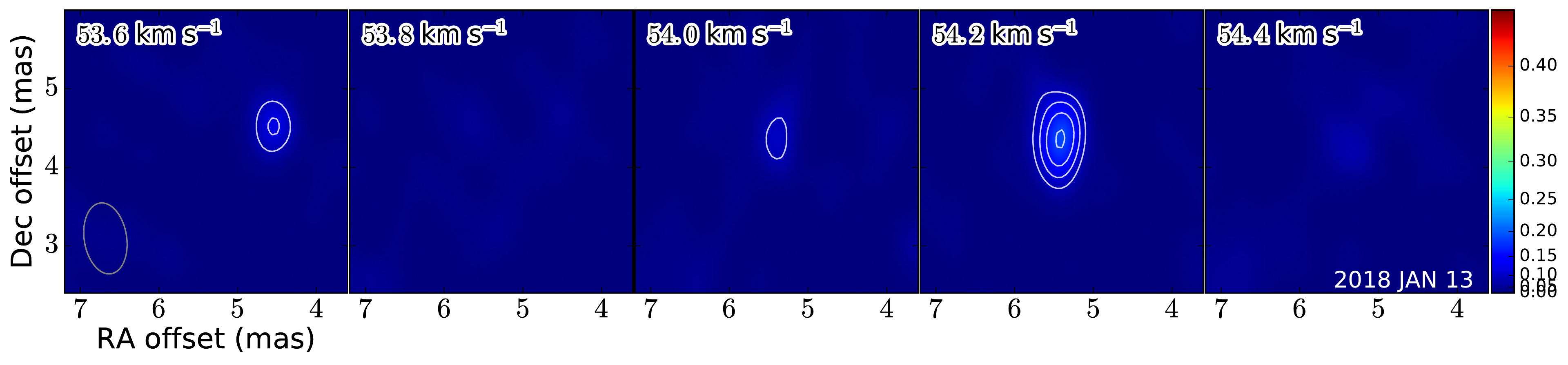}} 
 {\includegraphics[width=0.8\textwidth,angle=0]{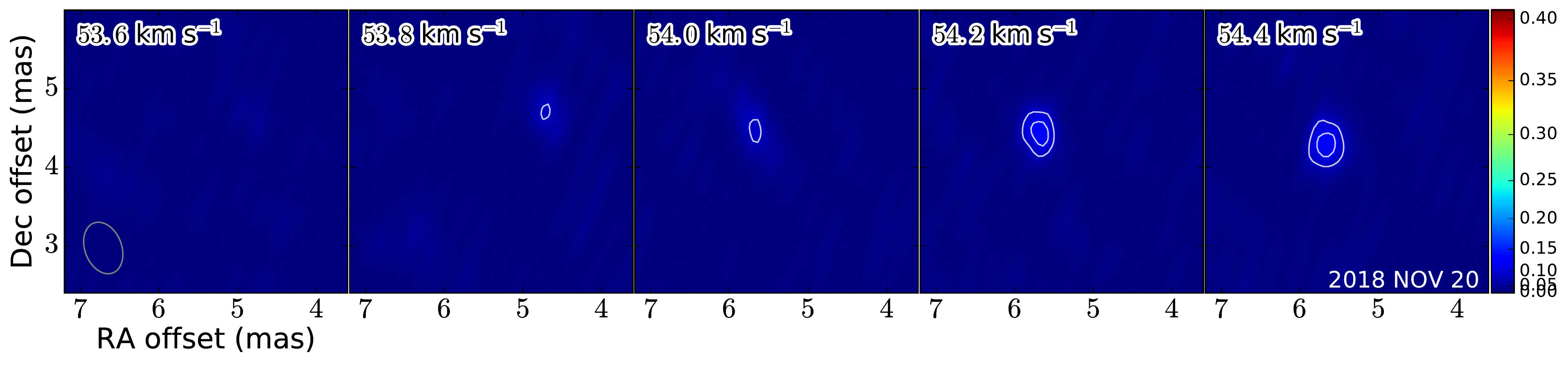}} 
 {\includegraphics[width=0.8\textwidth,angle=0]{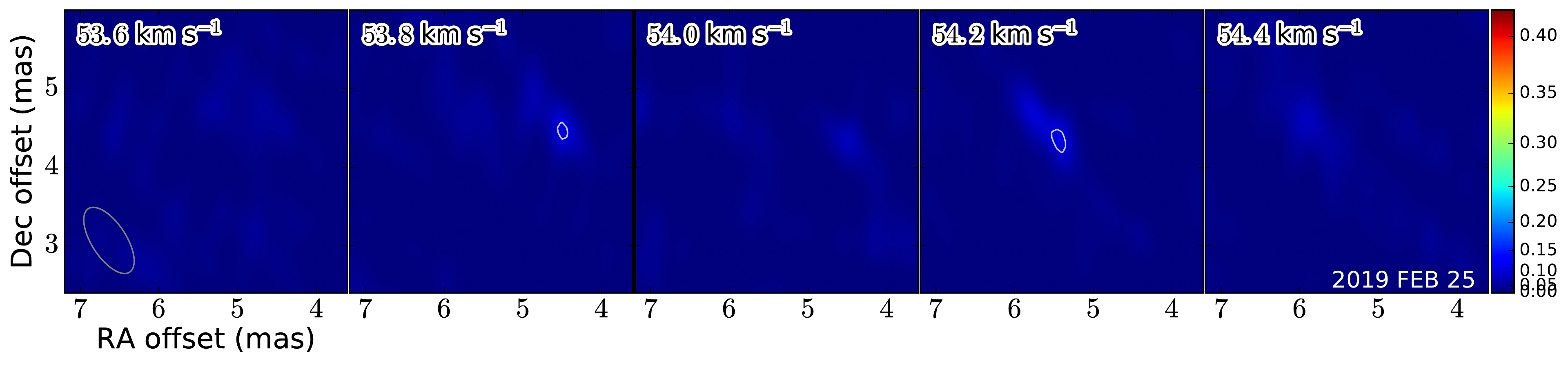}} 
 \end{center} 
\caption{Channel velocity maps of the SiO maser ($v=1$, $J=1-0$)
emission toward the north-east of the main spot at four epochs (see
dates in lower right corner). The offsets are relative to the peak
position of the strongest spot in each epoch. The contour levels
are at $5\sigma$, $7\sigma$, $9\sigma$, and $11\sigma$. }               
\label{fig:e1b} 
\end{figure*}

\begin{figure*}[!bht] 
\begin{center} 
 {\includegraphics[width=0.9\textwidth,angle=0]{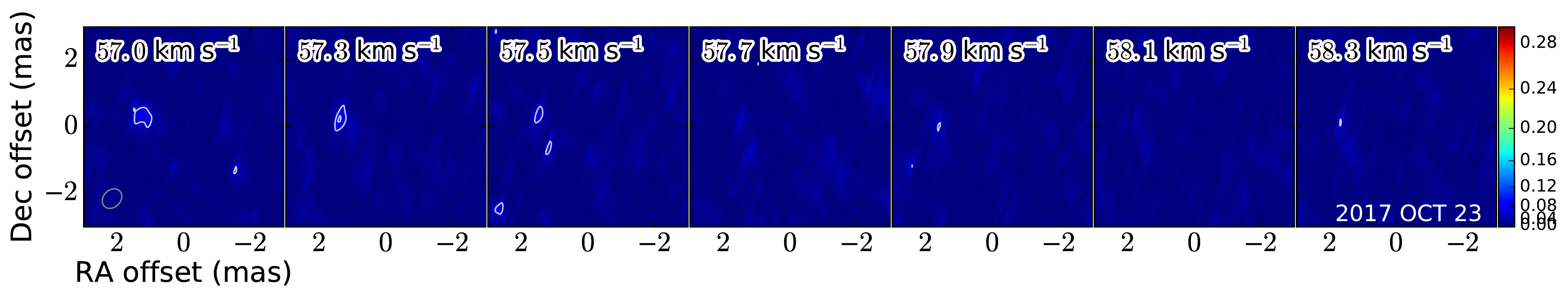}} 
 {\includegraphics[width=0.9\textwidth,angle=0]{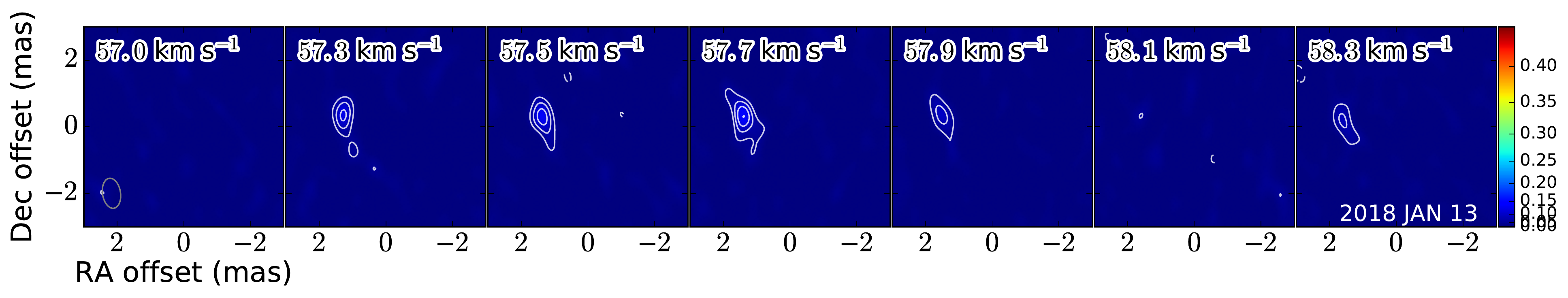}} 
 \end{center} 
\caption{Channel velocity maps of the SiO maser ($v=1$, $J=1-0$)
emission from the third spot. The offsets are relative to the peak
position of the strongest spot in each epoch. The contour levels
are at $3\sigma$, $5\sigma$, $7\sigma$, and $9\sigma$. }                
\label{fig:spot3} 
\end{figure*}

\begin{figure*}[!bht] 
\begin{center} 
 {\includegraphics[width=0.5\textwidth,angle=0]{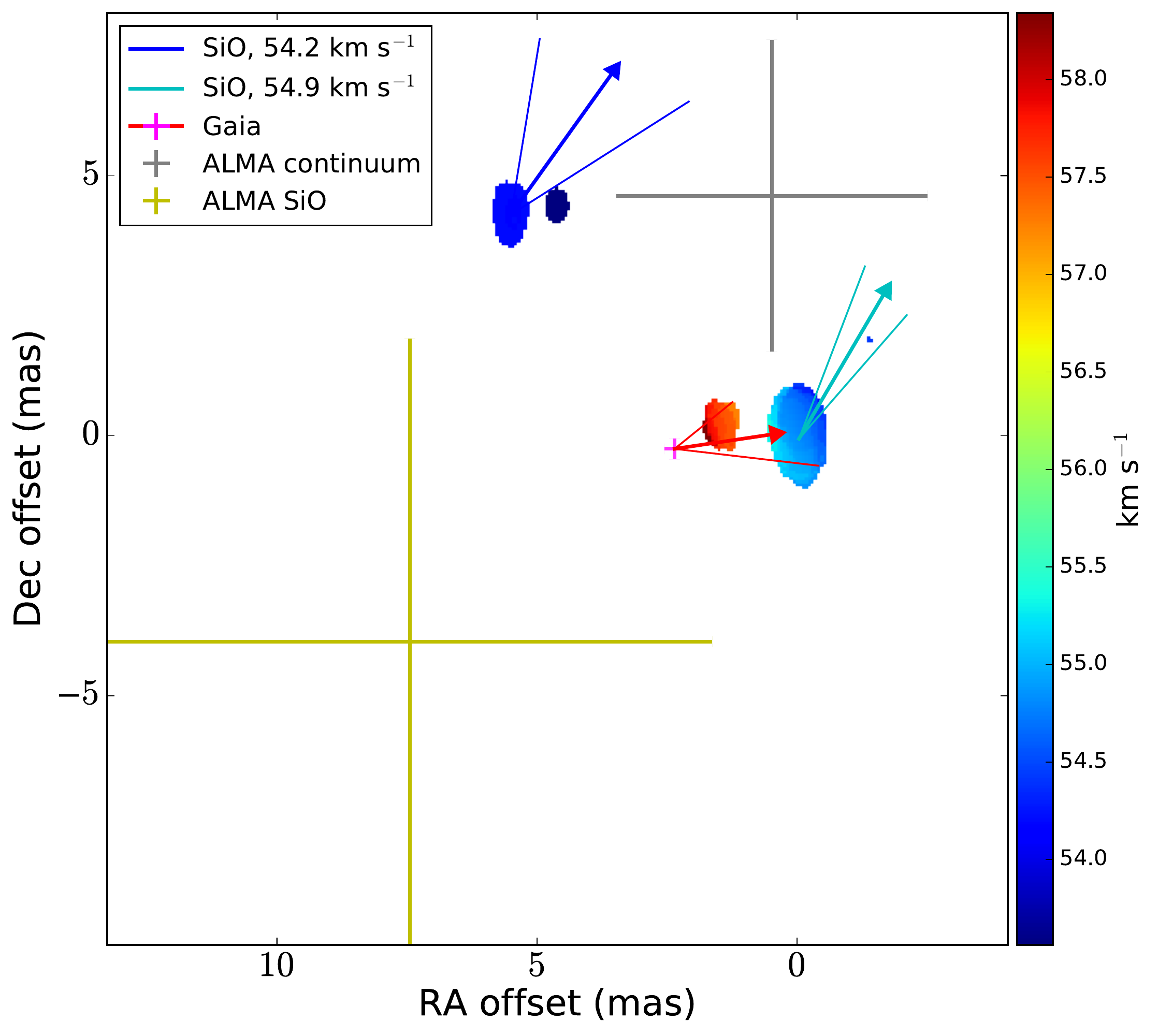}} 
 \end{center} 
\caption{Location of the spots detected on 13 January 2018 for the
$v=1$, $J=1-0$ SiO maser transition.                                    
The  proper motion vectors of the two strongest spots (at 54.9 and
54.2 km~s$^{-1}$) are indicated by the cyan and blue arrows, respectively.
A third spot is detected at $\sim$58~km~s$^{-1}$ and located at $\sim$1~mas
to the east of the strongest spot. The location of the Gaia source
and its proper motion is indicated by the pink cross and the red
arrow, respectively. The lines associated to the arrows indicate
the proper motion errors. The grey and yellow crosses mark the positions
and uncertainties of the continuum and SiO extended emission peaks
detected in the ALMA maps, respectively.}                               
\label{fig:spots} 
\end{figure*}

\begin{figure*}[!bht] 
\begin{center} 
 {\includegraphics[width=0.45\textwidth,angle=0]{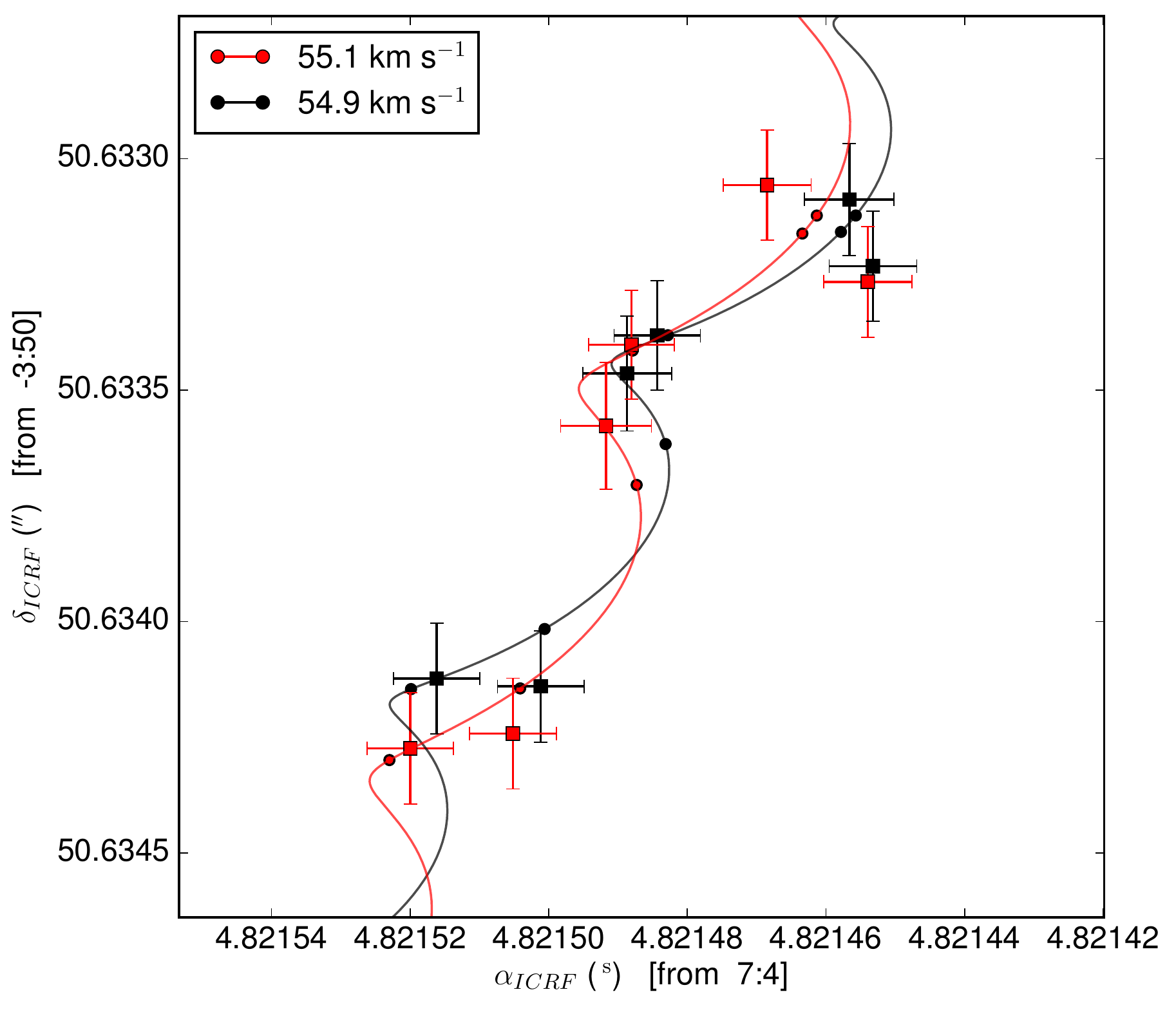}} 
 \hspace{0.7cm} 
  {\includegraphics[width=0.45\textwidth,angle=0]{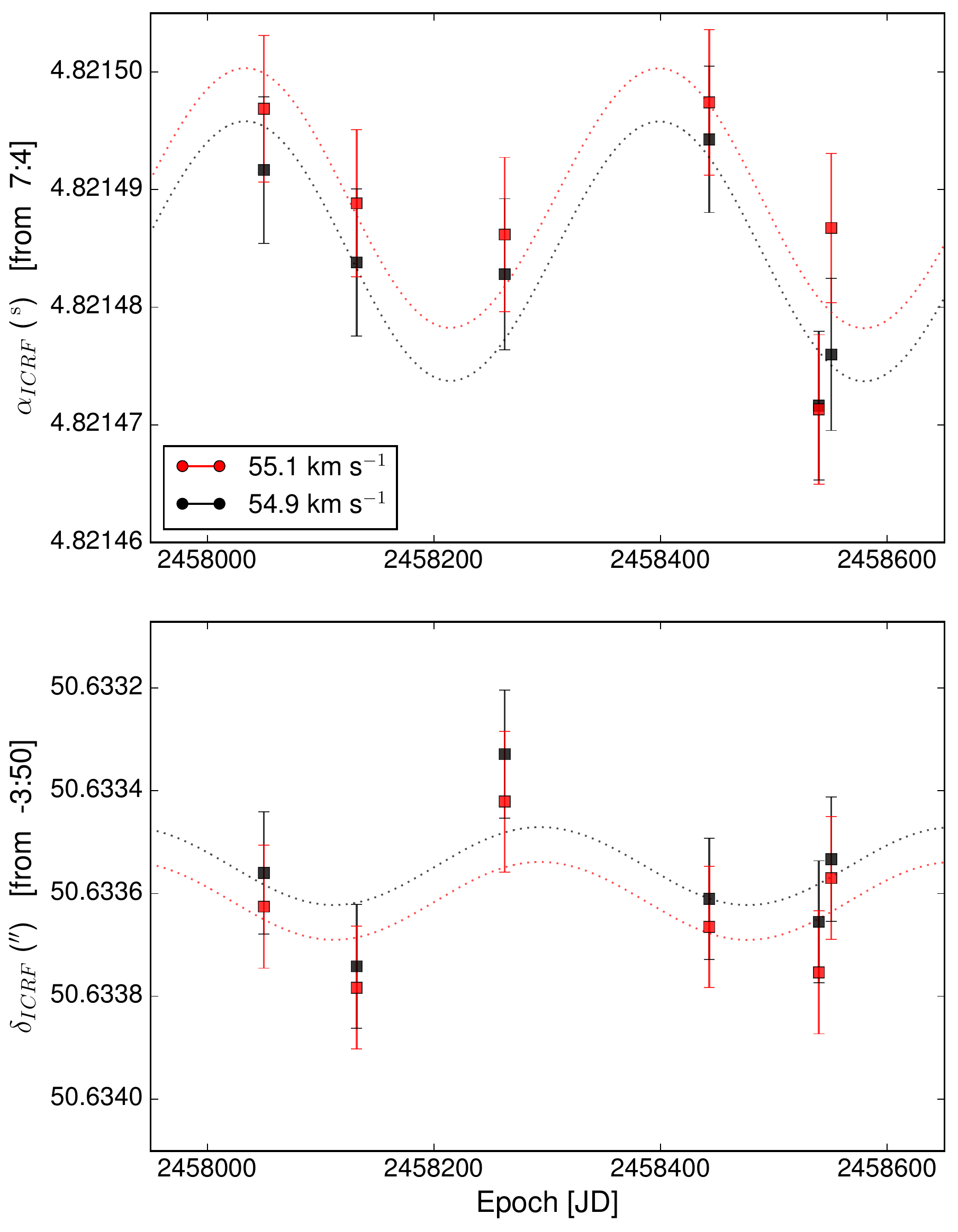}} 
 \end{center} 
\caption{{\it Left:} Measured VLBA sky positions of V838~Mon maser
spots in two contiguous channels (black and red squares). The line
corresponds to the best fit to the data while the filled circles
are the expected positions from this fit at the observed epochs.
{\it Rigth:}  right ascension (upper panel) and declination (lower
panel) of the maser spots as a function of time after subtracting
the fitted proper motions to see only the effect of the  parallax. }    
\label{fig:fit} 
\end{figure*} 
 
\begin{figure*}[!bht] 
\begin{center} 
\includegraphics[width=0.48\textwidth,angle=0]{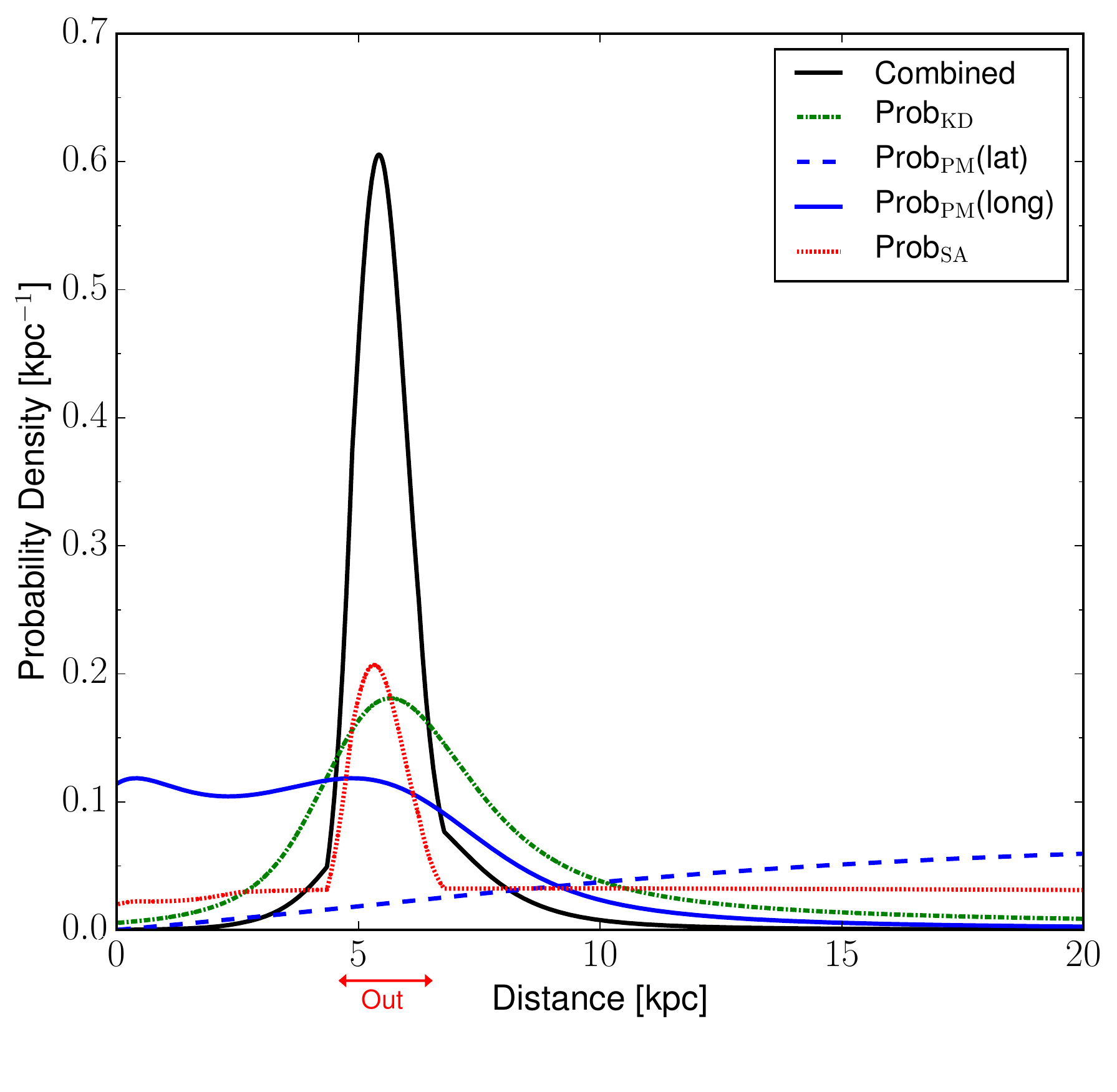}
\includegraphics[width=0.47\textwidth,angle=0,trim=0 -9mm 0 0]{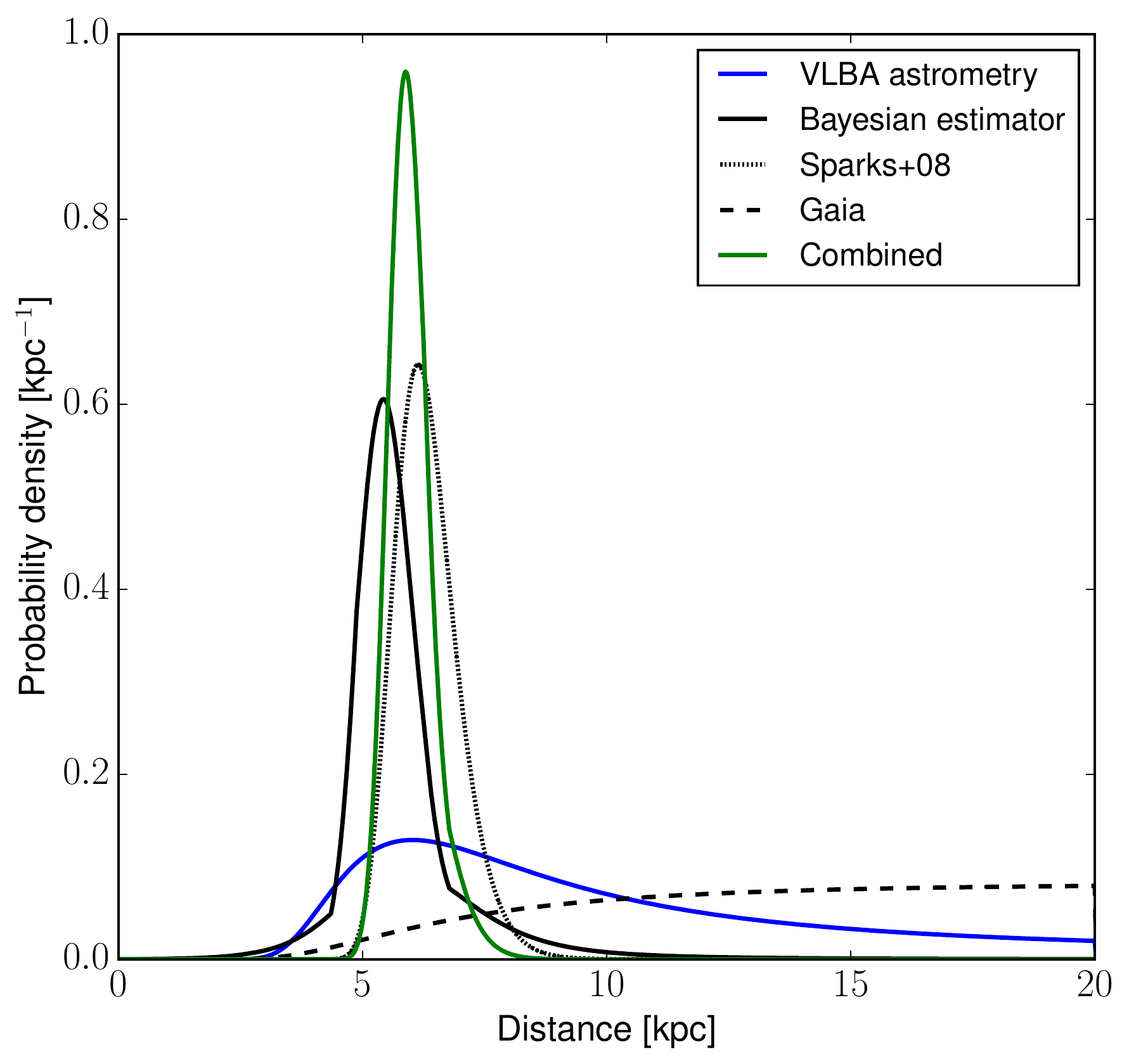}
  \end{center} 
\caption{{\it Left:} Normalized distance PDF for V838~Mon (black
line) which combines kinematic distance (green dashed line), Galactic
latitude and spiral arm model (red dashed line) and source's proper
motion in Galactic longitude (blue solid line) and Galactic latitude
(blue dashed line). 
     The red arrow marks the distance range for possible arm associations.
{\it Right:} Normalized distance PDF for the various measurements
of the V838~Mon parallax, including our VLBA astrometry, the Bayesian
estimator, Gaia-DR2 and the light echo.                                 
} 
\label{fig:pdf-all} 
\end{figure*}

\begin{figure*}[!bht] 
\begin{center} 
 {\includegraphics[width=0.5\textwidth,angle=0]{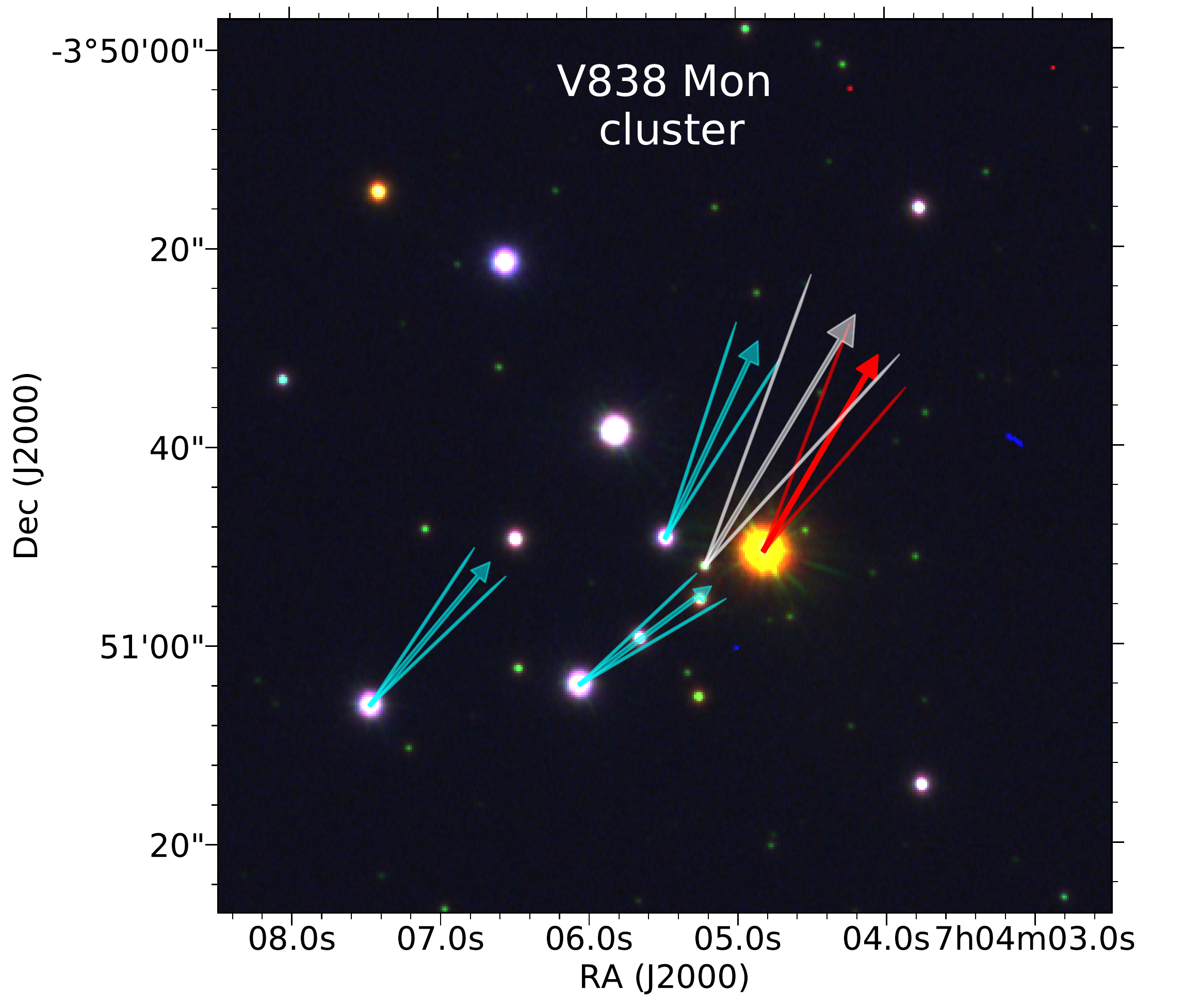}}
 \end{center} 
\caption{Proper motions of the identified members of the young open
cluster of V838 Mon. The arrows show the proper motions scaled by
a factor of 25\,000 and the associated lines indicate the errors
of the measurements. Red arrow corresponds to our VLBA measurement
for V838 Mon, blue arrows correspond to Gaia DR2 values for identified
cluster members, and the grey arrow is drawn for a new candidate
member of the cluster and represents its proper motion in the Gaia
DR2 catalog. The background image is a color rendition of images
obtained in 2015 with ESO--VST/OMEGACAM \citep{Drew}; red, green,
and blue colors correspond to $r$, $g$, and $u$ filters, respectively.}\label{fig-GaiaCluster}
\end{figure*}

\begin{figure*}[!bht] 
\begin{center} 
 {\includegraphics[width=0.49\textwidth,angle=0]{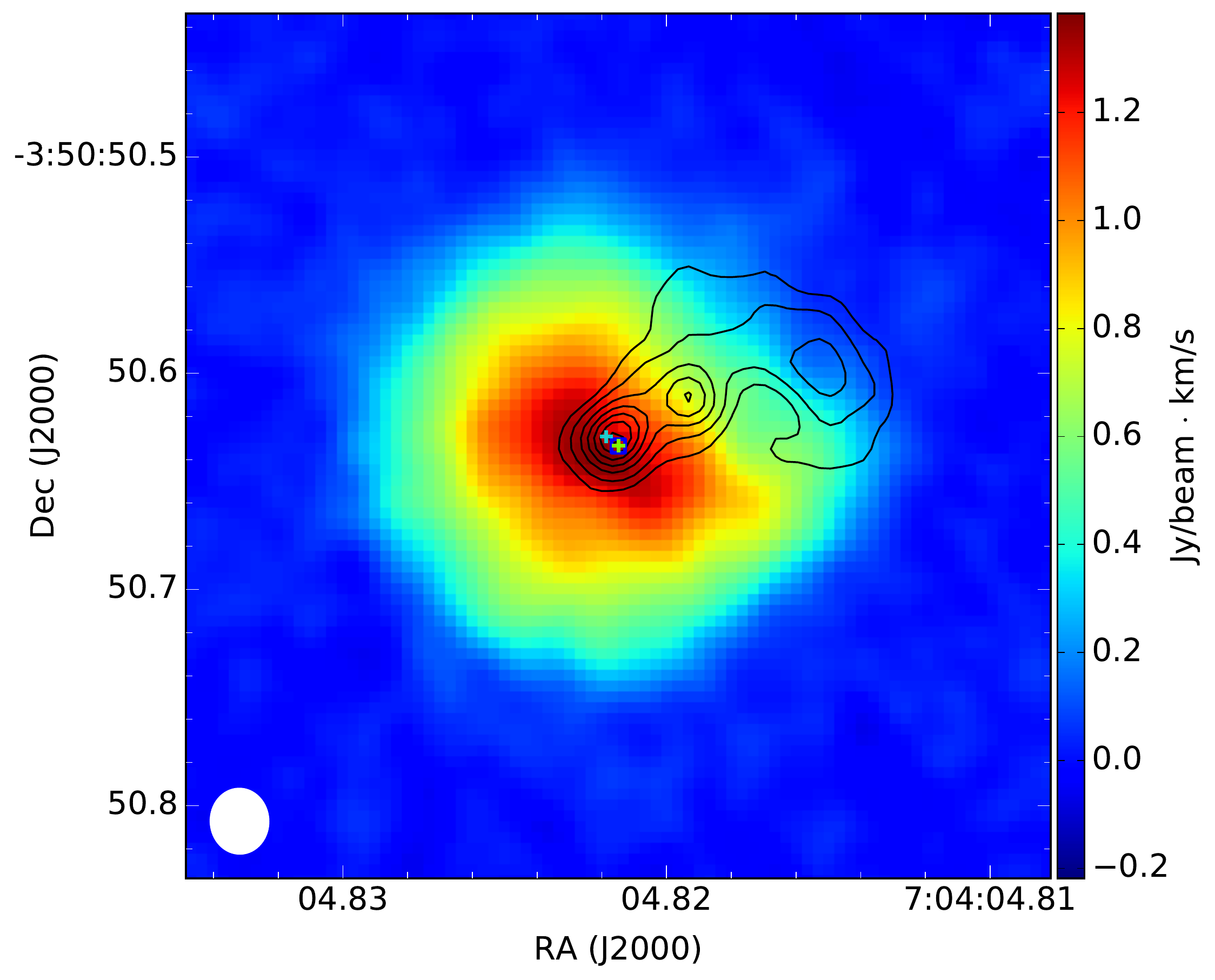}}
 {\includegraphics[width=0.49\textwidth,angle=0]{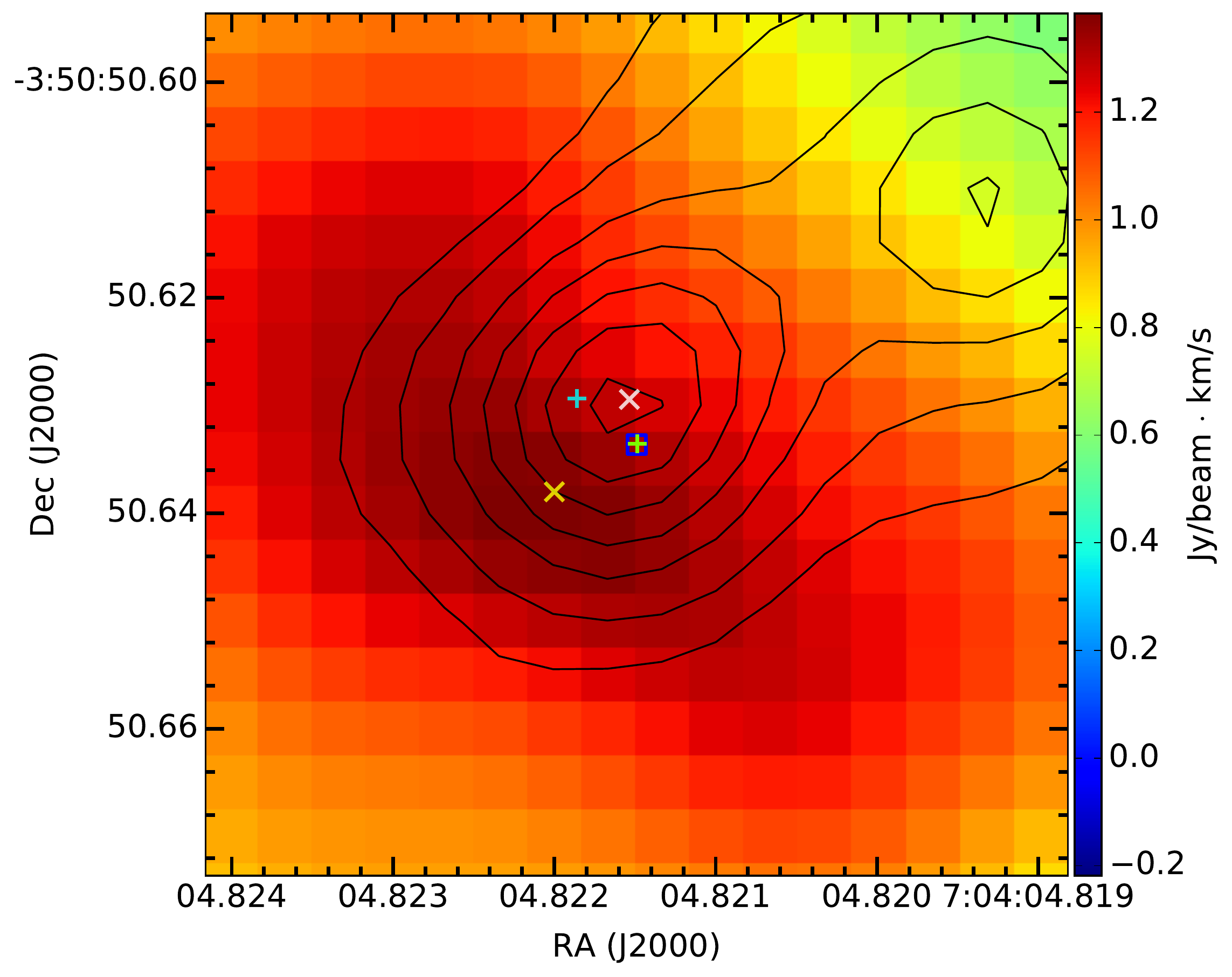}}
 \end{center} 
\caption{Moment zero map of the SiO molecular emission observed with
ALMA. The contours show the continuum emission at 10, 20, 35, 50,
65, 80 and 95\% of the peak emission. The positions of the SiO masers
detected with the VLBA are indicated by two the pluses and the square.
The right panel shows a zoom-in of the central part of the map. The
white and yellow crosses mark the continuum and SiO extended emission
peaks detected in the ALMA maps, respectively.}\label{fig:alma}         
\end{figure*} 

\begin{acknowledgements} 
The authors sincerely acknowledge the anonymous referee, whose comments
improved the manuscript. G.N.O.-L. acknowledges support from the
Alexander von Humboldt Foundation. TK acknowledges funding from grant
no 2018/30/E/ST9/00398 from the Polish National Science Centre. R.T.
acknowledges a support from grant 2017/27/B/ST9/01128 financed by
the Polish National Science Centre.                                     
 
The Long Baseline Observatory is a facility of the National Science
Foundation operated under cooperative agreement by Associated Universities,
Inc. The National Radio Astronomy Observatory is a facility of the
National Science Foundation operated under cooperative agreement
by Associated Universities, Inc.                                        
 
This work has made use of data from the European Space Agency (ESA)
mission {\it Gaia} (\url{https://www.cosmos.esa.int/gaia}), processed
by the {\it Gaia} Data Processing and Analysis Consortium (DPAC,
\url{https://www.cosmos.esa.int/web/gaia/dpac/consortium}). Funding
for the DPAC has been provided by national institutions, in particular
the institutions participating in the {\it Gaia} Multilateral Agreement.
This paper makes use of the following ALMA data: ADS/JAO.ALMA\#2018.1.00336.S.
ALMA is a partnership of ESO (representing its member states), NSF
(USA) and NINS (Japan), together with NRC (Canada), MOST and ASIAA
(Taiwan), and KASI (Republic of Korea), in cooperation with the Republic
of Chile. The Joint ALMA Observatory is operated by ESO, AUI/NRAO
and NAOJ. The National Radio Astronomy Observatory is a facility
of the National Science Foundation operated under cooperative agreement
by Associated Universities, Inc.                                        
Based on observations collected at the European Organisation for
Astronomical Research in the Southern Hemisphere under ESO programme
177.D-3023(F).                                                          
\end{acknowledgements}

 
\bibliographystyle{aa} 
\bibliography{ms.bib} 
 
\end{document}

%% file: observations.tex
\clearpage
\begin{table}[ht]
\caption{VLBA observed epochs}
\label{tab:obs} 
\centering 
\begin{tabular}{l l c c c }  
\hline\hline  
 ID BM464 & Observation Date   & JD & Beam  Size          & rms               \\
          &                    &    & (mas$\times$mas; deg)   & (mJy/beam)   \\
\hline 
A & 2017 Oct 23 & 2458049.99042 & 0.65$\times$0.51;~-45.0$^{\rm o}$   & 20 \\
B & 2018 Jan 13 &  2458131.76652 & 0.91$\times$0.54;~~~~9.0$^{\rm o}$& 17 \\
C & 2018 May 23 & 2458262.40884 & 0.74$\times$0.23;~-16.2$^{\rm o}$ & 17 \\
D & 2018 Aug 27 &  2458358.64671 & 0.91$\times$0.43;~~28.4$^{\rm o}$ & 27 \\
E & 2018 Nov 20 &  2458442.91462 & 0.68$\times$0.47;~~20.8$^{\rm o}$ & 17 \\
F & 2019 Feb 25 &  2458539.64977 & 0.96$\times$0.44;~~33.0$^{\rm o}$ & 21 \\
F1 & 2019 Mar 07 & 2458550.61973 & 0.75$\times$0.42;~16.6$^{\rm o}$ & 20 \\
\hline 
\end{tabular}
\end{table}

%% file: vlba_positions.tex
\begin{table}[ht]
\caption{Source properties}
\label{tab:jmfit} 
\centering 
\begin{tabular}{c c c c c c c c c}  
\hline\hline  
 JD    &  $\alpha$  &  $\sigma_\alpha$& $\delta$ & $\sigma_\delta$  & Peak Flux &  Integrated Flux    \\
       & 07$^{\rm h}$~04$^{\rm m}$ & (s) & $-03^{\rm o}~50{'}$  & (arcsec) &  (Jy/beam)      &  (Jy)  \\
\hline 
Maser V$_{\rm LSR}=55.1~{\rm km~s}^{-1}$ \\
2458049.99042 &  4.8215200 & 9.5e-07  & 50.634274 & 3.0e-05  & 0.16 $\pm$ 0.02 & 0.24 $\pm$ 0.04  \\ 
2458131.76652 &  4.8215052 & 1.0e-06  & 50.634242 & 2.8e-05  & 0.22 $\pm$ 0.02 & 0.24 $\pm$ 0.04  \\ 
2458262.40884 &  4.8214917 & 2.2e-06  & 50.633577 & 7.3e-05  & 0.10 $\pm$ 0.02 & 0.29 $\pm$ 0.06  \\ 
2458442.91462 &  4.8214881 & 6.3e-07  & 50.633402 & 2.0e-05  & 0.22 $\pm$ 0.02 & 0.32 $\pm$ 0.04  \\ 
2458539.64977 &  4.8214540 & 1.5e-06  & 50.633266 & 2.9e-05  & 0.26 $\pm$ 0.02 & 0.30 $\pm$ 0.04  \\ 
2458550.61973 &  4.8214685 & 1.5e-06  & 50.633057 & 2.7e-05  & 0.22 $\pm$ 0.02 & 0.25 $\pm$ 0.04  \\
\hline 
Maser V$_{\rm LSR}=54.9~{\rm km~s}^{-1}$ \\
2458049.99042 &  4.8215162 & 8.9e-07  & 50.634123 & 2.7e-05  & 0.18 $\pm$ 0.02 & 0.29 $\pm$ 0.05  \\ 
2458131.76652 &  4.8215011 & 1.0e-06  & 50.634140 & 3.2e-05  & 0.22 $\pm$ 0.02 & 0.25 $\pm$ 0.04  \\ 
2458262.40884 &  4.8214887 & 1.8e-06  & 50.633464 & 4.5e-05  & 0.12 $\pm$ 0.02 & 0.23 $\pm$ 0.05  \\ 
2458442.91462 &  4.8214843 & 7.9e-07  & 50.633382 & 2.2e-05  & 0.22 $\pm$ 0.02 & 0.36 $\pm$ 0.05  \\ 
2458539.64977 &  4.8214532 & 1.4e-06  & 50.633232 & 2.5e-05  & 0.28 $\pm$ 0.02 & 0.35 $\pm$ 0.05  \\ 
2458550.61973 &  4.8214566 & 1.9e-06  & 50.633088 & 3.4e-05  & 0.19 $\pm$ 0.02 & 0.23 $\pm$ 0.04  \\ 
\hline 
\end{tabular}
\end{table}

%% file: vlba_astro.tex
\begin{table}[ht]
\caption{Astrometric parameters from VLBA}
\label{tab:astro} 
\centering 
\begin{tabular}{c c  c c c c c}  
\hline\hline  
Maser V$_{\rm LSR}$ &  $\alpha_0$ & $\delta_0$ &  $\mu_\alpha \cos \delta$ & $\mu_\delta$ & $\varpi$ & D\tablefootmark{a} \\
(km~s$^{-1}$)       &  07$^{\rm h}$~04$^{\rm m}$ & $-03^{\rm o}~50{'}$ & (mas/yr) &  (mas/yr)  & (mas) & (kpc) \\
 \hline
55.1 &  $4\rlap.{^{\rm s}}82149(7)$ 
& $50\rlap.{''}6336(1)$
& -0.451 $\pm$ 0.079  &   0.847 $\pm$ 0.092 & \multirow{2}{*}{0.166 $\pm$ 0.060} & \multirow{2}{*}{$6.0^{+3.4}_{-1.6}$}\\
54.9 &  $4\rlap.{^{\rm s}}82148(7)$ 
& $50\rlap.{''}6335(1)$ 
& -0.479 $\pm$  0.079  &  0.736 $\pm$ 0.092    \\
Average & -- & -- & -0.465 $\pm$ 0.056 & 0.791 $\pm$ 0.065&--&-- \\
\hline
54.2 & 4.82186(2) & 50.6293(1) & -0.553 $\pm$ 0.339 & 0.771 $\pm$ 0.167  & -- & -- \\
\hline
\end{tabular}
\tablefoot{Numbers in parenthesis give position errors in units of the last significant digits.}
\tablefoottext{a}{Distance inferred from the inversion of the parallax. }
\end{table}
\clearpage

%% file: parallaxes_all.tex
\begin{table*}[ht]
\caption{Parallax measurements}
\label{tab:prlx} 
\centering 
\begin{tabular}{c c c c }  
\hline\hline  
 $\varpi$ [mas]              && Instrument & Reference   \\
\hline 
 $-0.001\pm0.105$ && {\it Gaia} astrometry    & 1, 2 \\  
 $0.166\pm0.060$  && VLBA astrometry        & 3  \\
 \hline
  $\varpi$ [mas]              & Distance [kpc] & Method & Reference   \\
 \hline
$0.164\pm0.016$ & $6.1\pm0.6$            & light-echo polarimetry & 4     \\
$0.182\pm0.019$ & $5.5\pm0.6$ &  Distance estimator & 3 \\ 
$0.181\pm0.018$ & $5.5\pm0.6$ & Distance estimator+VLBA & 3 \\
$0.169\pm0.011$ & $5.9\pm0.4$ & Distance estimator+VLBA+light echo+{\it Gaia} & 3 \\
\hline 
\end{tabular}
\tablebib{(1)~\citet{Gaia2018}; (2)~\citet{Bailer-Jones2018}; (3)~This work; (4)~\citet{Sparks2008}. 
}
\end{table*}